\documentclass[paper]{JHEP3}
\pdfoutput=1
\usepackage{amsmath,amssymb,amsthm,amscd,graphicx}
\input epsf.sty

\theoremstyle{definition}

\newcommand{\CA}{{\cal A}}
\newcommand{\CB}{{\cal B}}
\newcommand{\CC}{{\cal C}}

\newcommand{\CF}{{\cal F}}

\newcommand{\CH}{{\cal H}}

\newcommand{\CJ}{{\cal J}}

\newcommand{\CL}{{\cal L}}

\newcommand{\CN}{{\cal N}}
\newcommand{\CO}{{\cal O}}

\newcommand{\CT}{{\cal T}}

\def\IZ{{\mathbb Z}}
\def\IR{{\mathbb R}}

\def\IP{{\mathbb P}}

\def\IS{{\mathbb S}}

\def\l{\ell}
\newcommand{\tr}{{\rm Tr}}
\newcommand{\re}{{\rm e}}
\newcommand{\ri}{{\rm i}}
\newcommand{\rd}{{\rm d}}

\newcommand{\be}{\begin{equation}}
\newcommand{\ee}{\end{equation}}
\newcommand{\ba}{\begin{aligned}}
\newcommand{\ea}{\end{aligned}}
\newcommand{\ben}{\begin{eqnarray}\displaystyle}
\newcommand{\een}{\end{eqnarray}}

\newcommand{\sectiono}[1]{\section{#1}\setcounter{equation}{0}}

\newdimen\tableauside\tableauside=1.0ex
\newdimen\tableaurule\tableaurule=0.4pt
\newdimen\tableaustep
\def\phantomhrule#1{\hbox{\vbox to0pt{\hrule height\tableaurule width#1\vss}}}
\def\phantomvrule#1{\vbox{\hbox to0pt{\vrule width\tableaurule height#1\hss}}}
\def\sqr{\vbox{%
  \phantomhrule\tableaustep
  \hbox{\phantomvrule\tableaustep\kern\tableaustep\phantomvrule\tableaustep}%
  \hbox{\vbox{\phantomhrule\tableauside}\kern-\tableaurule}}}
\def\squares#1{\hbox{\count0=#1\noindent\loop\sqr
  \advance\count0 by-1 \ifnum\count0>0\repeat}}
\def\tableau#1{\vcenter{\offinterlineskip
  \tableaustep=\tableauside\advance\tableaustep by-\tableaurule
  \kern\normallineskip\hbox
    {\kern\normallineskip\vbox
      {\gettableau#1 0 }%
     \kern\normallineskip\kern\tableaurule}%
  \kern\normallineskip\kern\tableaurule}}
\def\gettableau#1{\ifnum#1=0\let\next=\null\else
\squares{#1}\let\next=\gettableau\fi\next}

\newcommand{\figref}[1]{Fig.~\protect\ref{#1}}

\title{Exact results in $\CN=8$ Chern-Simons-matter theories and quantum geometry}

\author{
Santiago Codesido, Alba Grassi and Marcos Mari\~no
\\
D\'epartement de Physique Th\'eorique et Section de Math\'ematiques,\\
Universit\'e de Gen\`eve, Gen\`eve, CH-1211 Switzerland\\
\\
\email{santiago.codesido@unige.ch, alba.grassi@unige.ch, marcos.marino@unige.ch}
}

\abstract{We show that, in ABJ(M) theories with $\CN=8$ supersymmetry, the non-perturbative sector of the partition function on the three-sphere 
simplifies drastically. Due to this simplification, we are able to write closed form expressions for the grand potential of these theories, which determines 
the full large $N$ asymptotics. Moreover, we find explicit formulae for the generating functionals of their partition functions, for all values of the rank $N$ of the 
gauge group: they involve Jacobi theta functions on the spectral curve associated to the planar limit. Exact quantization conditions for the spectral problem of the 
Fermi gas are then obtained from the vanishing of the theta function. We also show that the partition function, 
as a function of $N$, can be extended in a natural way to an 
entire function on the full complex plane, and we explore some 
possible consequences of this fact for the quantum geometry of M-theory and for putative de Sitter extensions.}

\begin{document}

\sectiono{Introduction}

In the last years, the techniques of localization have led to many results in supersymmetric quantum field theories. After the use of localization, the path integral 
is typically reduced to a matrix model. This is indeed an enormous simplification, but in many cases one needs to go further and extract analytic information from the resulting expression. 
For example, in applications to AdS/CFT, one is typically interested in the large $N$ behavior of the matrix model, which is indeed a non-trivial problem. 

The localization of supersymmetric Chern--Simons--matter theories was initiated in \cite{kwy}, where the partition function on the three-sphere was calculated 
for theories with a least $\CN=3$ supersymmetry. The corresponding matrix models have been analyzed in detail by using different techniques. 
In the case of ABJ(M) theory \cite{abjm,abj}, the partition function 
has been computed both at finite $N$ and at large $N$. In particular, the full large $N$ asymptotics of the ABJM matrix integral, including non-perturbative 
corrections, has been found explicitly in a series of papers \cite{mpabjm,dmp,mp,hmo2,hmo3, cm,hmmo,km}. The results have been also generalized to ABJ theory 
\cite{awhs, honda,matsumori,hondao,kallen}. In both cases, the large $N$ asymptotics is expressed in terms of a subsidiary theory, 
the (refined) topological string on a non-compact 
Calabi--Yau known as local $\IP^1 \times \IP^1$. It involves the 
topological closed string amplitudes at all genus, as well as the all-genus expansion of the Nekrasov--Shatashvili (NS) limit \cite{ns} of the refined theory. 

ABJ(M) theory has generically $\CN=6$ supersymmetry, but it is well known that in some cases there is an enhancement to $\CN=8$ supersymmetry. This happens for ABJM theory 
with Chern--Simons level $k=1,2$ \cite{bkk, gr,notes, kapustin}, and also for ABJ theory, when the ranks of the nodes differ in one unity and $k=2$ \cite{kapustin2}. 
One would expect that, in those cases, the theory simplifies in a substantial way (see for example \cite{pufu} for a recent study of these theories in the context of the conformal bootstrap). 
However, such a simplification has not been put to fruition in the studies on localization of these theories.  
One good reason is that it is not visible in the 't Hooft expansion of \cite{dmp}, nor in the perturbative M-theory expansion studied in \cite{hkps,fhm}.  

In this paper we will show that, indeed, a radical simplification takes place in the expression for the partition function of 
$\CN=8$ ABJ(M) theories, but it does so in the non-perturbative sector. As we mentioned before, 
this sector involves, in the $\CN=6$ case, the (refined) topological string at all genus. It turns out that, in theories with enhanced $\CN=8$ supersymmetry, 
only the $g=0,1$ amplitudes contribute. This is somewhat similar 
to what happens for topological strings on manifolds with reduced holonomy. Therefore, 
from the point of view of the subsidiary topological string, $\CN=8$ theories are one-loop exact. Due to this simplification, 
we are able to write down closed form expressions for the grand potential of these theories, which determines the full large $N$ expansion. 
Moreover, and more surprisingly, 
we find generating functionals for the partition functions at finite $N$. 
These partition functions were calculated up to high $N$ in \cite{py,hmo,hmo2}, and they are particularly simple: they are polynomials in $1/\pi$ with rational 
coefficients. Due to this simple form, it was suspected that there should be generating functionals for these numbers. In this paper 
we present completely explicit expressions for these generating functionals. They turn out to involve in a crucial way the 
non-perturbative partition functions for spectral curves of \cite{bde,eynard,em}, which are 
written in terms of Jacobi theta functions. As a further spinoff of this result, we derive exact quantization conditions for the spectral problem associated to the 
Fermi gas of $\CN=8$ ABJ(M) theories. 
They are determined by the zeros of the relevant theta functions, and they fully agree with the conjectures put forward in \cite{km,kallen}. 

Armed with these results, we can address some interesting conceptual problems concerning the quantum geometry of M-theory. 
As it is already implicit in \cite{dmp}, the 't Hooft expansion of ABJM theory leads to a picture of stringy geometry in the type IIA superstring dual 
which is very similar to what was obtained for $\CN=2$ theories in two dimensions \cite{agm,aspinwall}: worldsheet instantons correct the point-particle limit, 
and this leads to two ``phases" in the moduli space of the 't Hooft parameter. In the type IIA dual, these phases correspond to a large distance or geometric phase, 
and to a short-distance or non-geometric phase. However, this result is intrinsically perturbative from the point of view of the genus expansion. The question which 
originally triggered this investigation was: how is this stringy geometry changed when we go to M-theory? In other words, what is the quantum 
geometry of the target space as we go beyond string perturbation theory? When we try to address this question in the context of the 
AdS/CFT correspondence, we have to take into account that the target space is naturally discretized. For example, in ABJM theory, the AdS/CFT dictionary tells us that 
\be
\label{ads-dictio}
\left( {L \over \ell_p} \right)^6 \approx k N, 
\ee
where $L$ is the radius of the M-theory background given by AdS$_4\times \IS^7/\IZ_k$. Although this relationship is in principle only valid at large $N$ 
(and indeed it is known to have corrections \cite{bh,aho}), it indicates 
that, as we go to small distances in the target space, we face the intrinsic discreteness of $N$. However, we show in this paper that the partition function of ABJ(M) theory can be 
naturally promoted to an {\it entire} function of $N$ in the complex plane (for fixed $k$). This is in contrast to the results in the 't Hooft expansion, 
where the genus $g$ free energies have branch cuts which lead to the non-trivial 
phase structure mentioned above. Therefore, based on this analytic continuation of the M-theory partition 
function to complex $N$, one can say that quantum corrections ``erase" the non-trivial analytic structure of the 
semiclassical $1/N$ expansion. This is conceptually 
similar to what was found in \cite{mmss}, albeit in a different context. Our analytic continuation makes it even 
possible to consider the theory at negative $N$, where the theory might have an interpretation in terms of quantum gravity in de Sitter space. 

The organization of this paper is as follows. In section 2, we review the existing relevant results on the localization of the ABJ(M) partition function on the three-sphere, and the 
conjectural answer for the series of non-perturbative effects. In section 3 we show that these effects simplify drastically in the maximally supersymmetric cases, and we write down 
explicit expressions for the modified grand potential which determine completely the full large $N$ expansion of the partition function. 
In section 4 we use the results of section 3 and the connection to the formalism developed in \cite{bde,eynard,em} to write down explicit 
generating functionals for the partition functions. In section 5 we find the exact quantization conditions determining the spectrum of the Fermi gas in $\CN=8$ theories. In section 6, we 
discuss the implications of our results for the understanding of the quantum geometry of M-theory, and for possible de Sitter continuations. Finally, in section 7 we conclude and list some 
prospects for future research. The Appendix A summarizes results on the special geometry of local $\IP^1 \times \IP^1$ which are used throughout this paper, while 
Appendix B lists some results on the Jacobi 
theta functions which are used in section 4. 

\sectiono{Localization and non-perturbative effects}

The partition function of ABJ(M) theory on $\IS^3$ has been reduced, by localization, to a matrix integral of the form \cite{kwy}
 (see \cite{mmrev} for a review and a list of references):
\be
\label{kapmm}
\ba
&Z(N_1, N_2, k)\\
&={\ri^{-\frac{1}{2}(N_1^2-N_2^2)}\over N_1! N_2!} \int \prod_{i=1}^{N_1}{ \rd \mu_i  \over 2\pi} \prod_{j=1}^{N_2} {\rd \nu_j \over 2\pi}
 {\prod_{i<j} \left( 2 \sinh \left( {\mu_i -\mu_j \over 2}\right) \right)^2 \left(2 \sinh \left( {\nu_i -\nu_j \over 2}\right) \right)^2 \over 
\prod_{i,j}  \left(2 \cosh \left( {\mu_i -\nu_j \over 2}\right) \right)^2} \re^{-{\ri k \over 4 \pi}\left(  \sum_i \mu_i^2 -\sum_j \nu_j^2\right)}. 
\ea
\ee
The ABJM case corresponds to $N_1=N_2=N$. An alternative formulation of the theory, known as the Fermi gas approach, has been proposed in \cite{mp}. Let us first consider 
the ABJM case, where $N_1=N_2=N$, and let us denote the resulting ABJM partition function as $Z(N,k)$. One finds \cite{kwy-ids,mp}, 
\be
\label{fgasform}
Z(N,k)={1 \over N!} \sum_{\sigma  \in S_N} (-1)^{\epsilon(\sigma)}  \int  {\rd ^N x \over (2 \pi k)^N} {1\over  \prod_{i} 2 \cosh\left(  {x_i  \over 2}  \right)
2 \cosh\left( {x_i - x_{\sigma(i)} \over 2 k} \right)}. 
\ee
Using standard results in Statistical Mechanics, one can identify (\ref{fgasform}) as the partition function of a one-dimensional ideal Fermi gas with density matrix
\be
\label{densitymat}
\rho(x_1, x_2)={1\over 2 \pi k} {1\over \left( 2 \cosh  {x_1 \over 2}  \right)^{1/2} }  {1\over \left( 2 \cosh {x_2  \over 2} \right)^{1/2} } {1\over 
2 \cosh\left( {x_1 - x_2\over 2 k} \right)}. 
\ee
By using the Cauchy identity, we can also rewrite (\ref{fgasform}) as \cite{kwy-ids,mp}
\be
\label{tanh-form}
Z (N,k)={1\over N!}  \int \prod_{i=1}^N {\rd x_i \over 4 \pi k}  {1\over 2 \cosh {x_i \over 2} } \prod_{i<j} \left( \tanh \left( {x_i - x_j \over 2 k } \right) \right)^2. 
\ee
The kernel (\ref{densitymat}) defines a positive-definite, Hilbert--Schmidt operator $\hat \rho$ through
\be
\langle x | \hat \rho |x '\rangle =\rho(x,x'). 
\ee
By Mercer's theorem (see for example \cite{bot}, chapter IV), this operator is also of trace class. Its eigenvalues lead to a discrete, infinite set of real energies $E_n$ via the spectral problem,
\be
\label{spectral}
\int \rho(x, x') \phi_n(x')\, \rd x = \re^{-E_n} \phi_n(x),  \qquad n \ge 0, 
\ee
where $\phi_n(x)$ are $L^2(\IR)$ functions. The trace class property guarantees that all the spectral traces are finite:
\be
Z_\ell =\tr \hat \rho^\ell= \sum_{n \ge 0} \re^{-\ell E_n} < \infty.
\ee

In the ABJ case, i.e. when $N_1\not=N_2$, the Fermi gas formulation of the theory has been worked out in \cite{awhs,honda,hondao} (a different approach to ABJ theory has been 
proposed in \cite{matsumori}). Let us denote $N_1=N$ and $N_2=N+M$, and let us introduce the simpler matrix integrals,
\be
\label{abj-alt}
Z(N,k; M)= {1\over N!}  \int \prod_{i=1}^N {\rd x_i \over 4 \pi k} V_M(x_i) \prod_{i<j} \left( \tanh \left( {x_i - x_j \over 2 k } \right) \right)^2, 
\ee
where
\be
V_M (x)={1\over \re^{x/2} + (-1)^M \re^{-x/2}} \prod_{s=-{M-1\over 2}}^{M-1\over 2} \tanh {x+ 2 \pi \ri s\over 2 k}. 
\ee
Note that this function is real, since both $s$ and $-s$ appear in the product. 
It turns out that the partition function (\ref{kapmm}), in the ABJ case, is given by (\ref{abj-alt}), times the partition function of pure 
Chern--Simons theory for group $U(M)$ at level $k$, and an additional phase factor. The Fermi gas 
formulation of (\ref{abj-alt}) leads to a density matrix given by the following kernel, 
\be
\rho(x_1, x_2)={1\over 4 \pi k}  { V_M^{1/2} (x_1) V_M^{1/2} (x_2)\over 
2 \cosh\left( {x_1 - x_2\over 2 k} \right)}.
\ee
It is again easy to see that this defines a positive-definite, trace class operator, and therefore a discrete, real spectrum of energies as in (\ref{spectral}). 

Although localization reduces the path integral of ABJ(M) theory to a matrix model, one still has to evaluate the resulting integrals. There are two possible 
approaches to this problem. First, one can try to compute the partition functions above for finite values of $N_1, N_2, k$. This can be done recursively by combining the Fermi gas approach 
with the results of \cite{zamo, tw}. This strategy was first proposed in \cite{py,hmo} and then pursued in \cite{hmo2, matsumori,hondao}. It does not lead though to closed formulae for the partition function 
as a function of $N$, $M$ and $k$. It rather 
leads to concrete expressions for $Z(N,k)$ or $Z(N, k;M)$ for low $N$ and fixed $k$, $M$. For example, for ABJM theory and $k=1$, one finds, for the very first values of $N$, 
\be
\label{zn1}
Z(1,1)= {1\over 4 }, \qquad Z(2,1)={1\over 16 \pi}, \qquad Z(3,1)={\pi-3  \over 64 \pi}.
\ee
In some cases, one can push the calculation to rather high values of $N$, although the procedure is indirect and requires 
computing various auxiliary functions.

\begin{figure}
\center
\includegraphics[height=6cm]{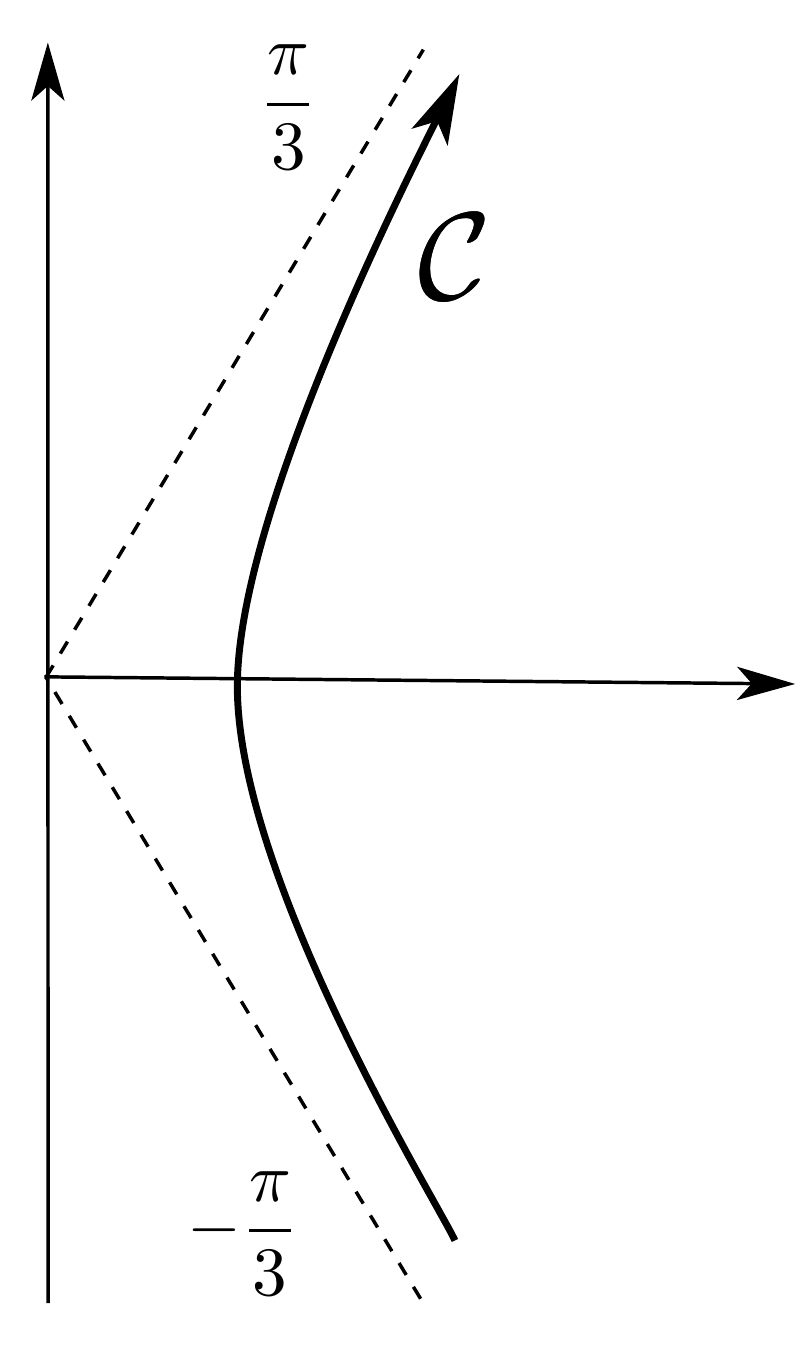}  
\caption{The standard contour in the complex plane of the chemical potential, defining the Airy function ${\rm Ai}$.}
\label{airy-c}
\end{figure}

The second approach, which is the most useful one in the applications to the AdS/CFT correspondence, involves a large $N$ analysis of the above integrals. 
In this case, having an explicit knowledge of the 
very first partition functions for $N$ finite is not very useful. However, one can use the Fermi gas 
approach to analyze the M-theory limit of the matrix integral, first considered in \cite{hkps}, in which $N$ is large and $k$ is finite. 
To state the results of this analysis, let us first focus on ABJM theory and let us introduce the grand canonical partition function, 
\be
\label{xips}
\Xi(\mu, k)=1+\sum_{N\ge 1} Z(N,k) \re^{N \mu}.
\ee
It can be easily shown \cite{mp} that this function can be represented as a Fredholm determinant for the integral operator with kernel (\ref{densitymat}), 
\be
\label{ximu}
\Xi (\mu, k)= \prod_{n\ge 0} \left(1+ \re^{\mu- E_n} \right). 
\ee
Since the operator $\hat \rho$ is of trace class, standard results (see for example \cite{simon}) 
show that (\ref{ximu}) is an analytic on the full complex plane of the fugacity\footnote{Usually, the fugacity is denoted by $z$. However, we will use a notation adapted to previous results in ABJM theory, and as we will show the fugacity will be identified with a parameter denoted by $\kappa$ in \cite{dmp}. In this paper, the letter $z$ will be 
used to denote the ``bare" coordinate parametrizing the moduli space of local $\IP^1\times \IP^1$.}, 
\be
\label{fugacity}
\kappa= \re^\mu. 
\ee
The conventional grand potential of the theory is defined as 
\be
\CJ(\mu, k) =\log \, \Xi (\mu, k). 
\ee
By using standard tools from Statistical Mechanics, one can show that \cite{mp}
\be
\label{cj-app}
\CJ (\mu,k)\approx {2 \mu^3\over 3 \pi^2 k} + \left( {k \over 24} + {1\over 3k} \right) \mu, \qquad \mu \gg 1. 
\ee
The partition function $Z(N,k)$ can then be obtained, for finite $N$, by extracting the $N$-th coefficient of the power series (\ref{xips}) through a contour integral around the origin, 
\be
\label{finite-n}
Z(N,k) ={1\over 2 \pi \ri} \oint \Xi(\kappa, k) \kappa^{-N-1} \rd \kappa = \int_{-\pi \ri}^{\pi \ri} {\rd \mu \over 2 \pi \ri} \re^{\CJ(\mu,k) - N \mu}. 
\ee
We can use the explicit result (\ref{cj-app}) to calculate $Z(N,k)$ in the saddle-point approximation, at large $N$ and fixed $k$ \cite{mp}. 
Since $\CJ(\mu, k)$ is a cubic polynomial, one deforms the contour in (\ref{finite-n}) to the standard contour $\CC$ for the Airy function, see \figref{airy-c}, and finds, 
\be
\label{app-sad}
Z(N,k) \approx \int_\CC  {\rd \mu \over 2 \pi \ri} \re^{\CJ(\mu,k) - N \mu} \approx C^{-1/3}(k) {\rm Ai} \left[ C^{-1/3}(k) \left( N- B(k) \right) \right], 
\ee
where
\be
\label{cb-ex}
C(k)= {2 \over  \pi^2 k}, \qquad B(k)= {k \over 24} + {1\over 3k}. 
\ee
Using the asymptotics of the Airy function at large $N$, one derives from (\ref{app-sad}) the $N^{3/2}$ behavior of M2-brane theories predicted in \cite{kt}. 
The result (\ref{app-sad}), expressing the partition function in terms of an Airy function, was first obtained in \cite{fhm} for ABJM theory by using the results of \cite{dmp,dmpnp}, and then 
rederived (and generalized) in \cite{mp} in this simpler language. It has been argued in \cite{ddg} that this result follows, to a large extent, from a localization calculation in supergravity. 

If we are interested in obtaining 
corrections to the large $N$ result (and in particular non-perturbative corrections at large $N$), it is important to be precise about the corrections to the saddle--point 
approximation (\ref{app-sad}). In \cite{hmo2} this problem was handled by introducing a different object: the modified (or ``naif") grand potential $J(\mu, k)$, which is {\it defined} by the 
equality 
\be
\label{naif}
Z(N,k)= \int_\CC  {\rd \mu \over 2 \pi \ri} \re^{J(\mu,k) - N \mu}. 
\ee
$J(\mu, k)$ is related to the conventional grand potential $\CJ(\mu,k)$ by \cite{hmo2}
\be
\label{naif-gp}
\re^{\CJ(\mu, k)} = \sum_{n \in \IZ} \re^{J(\mu+2 \pi \ri n , k)}. 
\ee
Indeed, if we plug this relationship in (\ref{finite-n}), we can use the sum over $n$ to extend the integration region in the second integral 
from $[-\pi \ri, \pi \ri]$ to the full imaginary axis. If we then deform the contour 
to $\CC$, we obtain (\ref{naif}). Notice that (\ref{naif-gp}) is manifestly invariant under 
\be
\mu \rightarrow \mu + 2 \pi \ri, 
\ee
which is indeed the case in view of the analyticity properties of $\Xi(\mu,k)$ as a function of the fugacity. 

Based on various works \cite{mp,dmp,dmpnp,mp,hmo,hmo2,cm,hmo3,hmmo,km}, a precise conjecture has emerged for the 
explicit structure of the modified grand potential $J(\mu,k)$. This conjecture 
has not been derived from first principles, although many of its ingredients have been verified analytically and numerically. As stated in its final form in \cite{hmmo}, the conjecture 
says that
\be
\label{gpmueff}
J(\mu, k)=J^{(\rm p)}(\mu_{\rm eff},k)+J^{\rm WS} (\mu_{\rm eff},k)+ \mu_{\rm eff}
\widetilde{J}_b(\mu_{\rm eff},k)+\widetilde{J}_c(\mu_{\rm eff},k).
\ee
In this expression, the perturbative piece is essentially the cubic polynomial in $\mu$ written down in (\ref{cj-app}), 
\be
J^{(\rm p)}(\mu,k)= {C(k) \over 3} \mu^3 + B(k) \mu + A(k), 
\ee
where the coefficients $C(k)$ and $B(k)$ are given in (\ref{cb-ex}). $A(k)$ is a function of $k$ which can be written in closed form as \cite{hanada, ho},
\be
\label{ak}
A(k)= \frac{2\zeta(3)}{\pi^2 k}\left(1-\frac{k^3}{16}\right)
+\frac{k^2}{\pi^2} \int_0^\infty \rd x \frac{x}{\re^{k x}-1}\log(1-\re^{-2x}).
\ee
The function $J^{\rm WS} (\mu,k)$, where ${\rm WS}$ stands for ``worldsheet instantons," is determined by topological string theory on the toric Calabi--Yau manifold 
known as local $\IP^1 \times \IP^1$ (in the so-called diagonal slice), and it can be written as 
\be
\label{gvone}
J^{\rm WS}(\mu, k)= \sum_{g\ge 0} \sum_{w,d \ge1} n^d_g \left( 2 \sin {2 \pi w \over k} \right)^{2g-2} {(-1)^{d w}  \over w}{\rm e}^{-{4 d w \mu \over k}}, 
\ee
where $n^d_g$ are the ``diagonal" Gopakumar--Vafa invariants \cite{gv} of local $\IP^1 \times \IP^1$ (see \cite{hmmo} for more details). The two functions $\widetilde{J}_b(\mu,k)$ and
$\widetilde{J}_c(\mu,k)$ take into account the membrane instanton contributions first identified in \cite{dmpnp,mp}. When expanded at large $\mu$, 
they are expressed in terms of two sets of coefficients $\widetilde{b}_\l(k)$, $\widetilde{c}_\l(k)$:
\begin{align}
 \widetilde{J}_b(\mu,k)=\sum_{\l=1}^\infty\widetilde{b}_\l(k)\re^{-2\l\mu}, \qquad \widetilde{J}_c(\mu,k)
=\sum_{\l=1}^\infty\widetilde{c}_\l(k)\re^{-2\l\mu}. 
\end{align}
The coefficients $\widetilde b_\ell(k)$ can be computed from the so-called refined topological string invariants $N^{d_1,d_2}_{j_L,j_R}$ of local $\IP^1 \times \IP^1$ \cite{ikv}, as
\be
\label{blj}
\widetilde{b}_\ell(k)=-\frac{\ell}{2\pi}\sum_{j_L,j_R}\sum_{\ell=dw}\sum_{d_1+d_2=d}N^{d_1,d_2}_{j_L,j_R}q^{\frac{w}{2}(d_1-d_2)}
\frac{\sin\frac{\pi kw}{2}(2j_L+1)\sin\frac{\pi kw}{2}(2j_R+1)}{w^2\sin^3\frac{\pi kw}{2}}. 
\ee
It was conjectured in \cite{hmo3} that one has the following relationship, 
\begin{align}
\widetilde{c}_\l(k)=- k^2 \frac{\partial}{\partial k} 
\left(\frac{\widetilde{b}_\l(k)}{2\l k}\right), 
\label{bcrel}
\end{align}
which relates $\widetilde{c}_\l(k)$ to $\widetilde{b}_\l(k)$, and we will assume it to be true in what follows. Finally, the ``effective" chemical potential $\mu_{\rm eff}$ is given by 
\be
\label{mueff-mu}
\mu_{\rm eff}= \mu + {1\over C(k)} \sum_{\ell=1}^\infty a_\ell(k) \re^{-2\ell \mu}, 
\ee
where $a_\ell(k)$ are the coefficients of the so-called quantum mirror map of local $\IP^1 \times \IP^1$ introduced in \cite{acdkv}. 
More details and relevant references on these quantities can be found 
in \cite{hmmo}. Note that, since this conjectural result on the non-perturbative structure of the ABJM partition function 
involves the modified grand potential $J(\mu, k)$, the original grand potential $\CJ (\mu, k)$ has 
played little r\^ole, except in the work \cite{morb}. 

All these results have been obtained for ABJM theory, but there are also similar results for ABJ theory. 
As in \cite{hondao}, we can focus on the matrix integral (\ref{abj-alt}), since the ABJ partition function can be 
obtained from it by known multiplicative factors. The grand canonical partition function is defined as 
\be
\Xi(\mu, k; M)=1+\sum_{N\ge 1} Z(N,k;M) \re^{N \mu},
\ee
and one can introduce a modified grand potential $J(\mu, k;M)$, similarly to what we did in (\ref{naif}) for ABJM theory. 
A conjectural form for this function has been proposed in \cite{matsumori,hondao}. 
It has a perturbative and a non-perturbative contribution, 
\be
J(\mu, k; M)=J^{(\rm p)}(\mu,k; M)+J^{(\rm np)}(\mu,k; M). 
\ee
The perturbative piece is again a cubic polynomial in $\mu$, 
\be 
J^{(\rm p)}(\mu,k; M)={2\mu^3\over 3 \pi^2 k}+B(k,M)\mu+A(k,M), 
\ee
with 
\be 
\label{abcons}
\ba B(k,M)&={1\over 3k}-{k\over 12}+{k \over 2}\left( {M \over k}-{1\over 2}\right)^2, \\
A(k,M)&= -\log{|Z_{\rm CS}^{(M)} (k)  |} +A(k). 
\ea
 \ee
 In (\ref{abcons}), $A(k)$ is given in (\ref{ak}), and $Z_{\rm CS}^{(M)} (k)$ is the partition function of Chern--Simons theory with gauge group $U(M)$ and level $k$, 
 \be
 Z_{\rm CS}^{(M)}(k)
=  k^{-\frac{M}{2}}  \prod_{s=1}^{M-1} \left( 2\sin{\frac{\pi s}{k} } \right)^{M-s}.
\ee
The non-perturbative piece has a particularly simple expression in the case $M=1$, 
which is the only one that will be considered in this paper. One has, from the general 
formulae in \cite{matsumori,hondao}, 
  \be 
  \label{abj-np}
  J(\mu, k; 1)=J^{(\rm p)}(\mu_{\rm eff},k; 1)+J^{\rm WS} (\mu_{\rm eff},k; 1)+\mu_{\rm eff}\widetilde{J}_b(\mu_{\rm eff}+{\ri \pi \over 2},k)+\widetilde{J}_c(\mu_{\rm eff}+{\ri \pi \over 2},k).
  \ee
The functions $\widetilde{J}_b(\mu,k)$, $\widetilde{J}_c(\mu,k)$ are the same ones which appear in (\ref{gpmueff}), 
 while the worldsheet instanton contribution is now given by 
 \be
 J^{\rm WS} (\mu,k;1) =\sum_{g\ge 0} \sum_{w,d \ge1} \sum_{d_1+d_2=d}(-1)^{dw} n^{d_1,d_2}_g \left( 2 \sin {2 \pi w \over k} \right)^{2g-2}{1 \over w} \re^{{2\pi \ri (d_1-d_2)w\over k}} {\rm e}^{-{4 d w \mu \over k}}. 
\ee
%
%
%
Finally, the expression for the effective potential $\mu_{\rm eff}$ appearing in (\ref{abj-np}) is explicitly known for any integer $k$, see \cite{hondao}. 

\sectiono{Exact results for $\CN=8$ ABJ(M) theories}

In this section we will evaluate the above expressions for the modified grand potential of ABJ(M) theory 
in the cases with maximal $\CN=8$ supersymmetry. This will allow us 
to write down a {\it closed} form expression for it. Mathematically, this is possible thanks to the equivalence between two 
different ways of writing the topological string free energies: the (refined) BPS representation of (\ref{gvone}), (\ref{blj}), and the B-model representation. 
The simplest case turns out to be ABJM theory with $k=2$. We will then start with this case. 

\subsection{ABJM with $k=2$} 

The first quantity we can evaluate in closed form is the ``effective" chemical potential $\mu_{\rm eff}$, as a function of $\mu$. 
This was done in \cite{hmo3} and the result is 
\be
\label{muk2}
\mu_{\rm eff}= \mu+2  \re^{-2\mu}  {}_4F_3 \left( 1,1, {3\over 2}, {3\over 2}; 2,2,2; - 16 \re^{-2 \mu} \right).
\ee
As we will see in this paper, all of our closed formula can be written in terms of the basic quantities appearing 
in the special geometry of diagonal, local $\IP^1 \times \IP^1$ (together with 
some ingredients of the off-diagonal theory). The relevant results are collected in Appendix A. In the so-called diagonal theory, the relevant moduli space of complex 
structures has complex dimension one and is parametrized by the ``bare" parameter $z$. The flat coordinate associated to this modulus 
is usually denoted by $t$, and the relation between them is given by the mirror map (\ref{mirror-map}). By comparing 
to (\ref{muk2}), we find that the relation between the parameters in ABJM theory and the parameters appearing in special geometry is 
 \be
 \label{dictioz}
z=\re^{-2 \mu}
 \ee
and
\be
\label{dictio2}
 t=2 \mu_{\rm eff}, 
 \ee
i.e. the chemical potential is related to the ``bare" parameter of mirror symmetry, while the ``effective" chemical potential $\mu_{\rm eff}$ is identified with 
the flat coordinate (up to a factor of $2$). This gives an interesting geometric interpretation to $\mu_{\rm eff}$, which 
was introduced in \cite{hmo3} to incorporate the contribution of bound states between membrane and worldsheet instantons. Prior to the inclusion of these bound states, 
and taking into account only the worldsheet instantons, one would naturally identify $t= 2 \mu$. Therefore, the effect of the bound states is to ``undo" the mirror map. 
 
Let us now study (\ref{gpmueff}) in the case $k=2$. The first thing to take into account is that all the summands in the r.h.s. (except of course the first one, which is just the 
perturbative piece) will have poles at $k=2$. However, 
it was shown in \cite{hmmo} that they all cancel at the end of the day (this is the HMO cancellation mechanism postulated in \cite{hmo2}). So we can just  
expand these summands around $k=2$, forget about the poles, and keep the finite part of the expansion (i.e. the coefficient of the $(k-2)^0$ term in the Laurent expansion around $k=2$). 

The first remarkable thing we find in this limit is that, for $k=2$, all terms with $g\ge 2$ in (\ref{gvone}) vanish. The $g=1$ contribution survives in the limit $k\rightarrow 2$, 
and we have to keep the finite part of $g=0$. 
An elementary calculation shows that the finite part of 
\be
 \left( 2 \sin {2 \pi w \over k} \right)^{-2} {\rm e}^{-{4 d w \mu \over k}}
\ee
is 
\be
{{\rm e}^{-2 d w \mu} \over 12 \pi^2 w^2} \left( 3+ \pi^2 w^2 + 6dw \mu + 6d^2 w^2 \mu^2\right). 
\ee
The finite piece of (\ref{gvone}) as $k\rightarrow 2$ is then, 
\be
 {\mu_{\rm eff}^2 \over 2 \pi^2} \partial^2_t F^{\rm inst}_0(t) - {\mu_{\rm eff} \over 2 \pi^2} \partial_t  F^{\rm inst}_0(t) +{1\over 4 \pi^2} F^{\rm inst}_0(t) + F^{\rm inst}_1(t), 
\ee
where
\be
F^{\rm inst}_0(t)= \sum_{w,d \ge1} n^d_0  {(-1)^{dw} \over w^3} \re^{-d w t}, \qquad F^{\rm inst}_1(t)= \sum_{w,d \ge1}  \left( {n^d_0 \over 12} + n^d_1\right) {(-1)^{wd} \over w}  \re^{-d w t}
\ee
are the genus zero and genus one free energies of the standard topological string. Here, and for the moment being, we only keep the instanton part of these free energies (i.e. we drop all the 
polynomial parts in $t$), and we use the dictionary (\ref{dictio2}). We conclude that for $k=2$ {\it only the $g=0,1$ topological string free energies 
contribute}. 

Let us now look at the membrane part. i.e. at the third and fourth summands in (\ref{gpmueff}). From the 
explicit expression (\ref{blj}), it is easy to see that the coefficient $\widetilde{b}_\ell(k)$ has the following behavior as $k\rightarrow 2$:
\be
\widetilde{b}_\ell(k)= {\widetilde{b}^{-1} _\ell \over k-2} + \widetilde{b}^{1}_\ell (k-2)+\cdots, 
\ee
therefore its finite part {\it vanishes} when $k\rightarrow 2$. The finite part of $\widetilde c_\ell (k)$ is then, from (\ref{bcrel}), 
\be
\widetilde{c}_\ell(k)=-{1\over \ell} \widetilde{b}^{1}_\ell+\cdots
\ee
From (\ref{blj}) we find the following expression,  
\be
 \widetilde{b}^{1}_\ell=\frac{\ell}{24}\sum_{j_L,j_R}\sum_{\ell=dw}\sum_{d_1+d_2=d}N^{d_1,d_2}_{j_L,j_R} {(-1)^{dw} \over w} m_L m_R \left( -3+ 3(d_1-d_2)^2 + m_L^2 + m_R^2 \right), 
 \ee
 where we have denoted
 \be
 m_L=2j_L+1, \qquad m_R = 2j_R+1. 
 \ee
As in the case of the worldsheet instanton contribution, we would like to express this quantity in terms of functions known in closed form. 
To do this, we compare the BPS expansion of the refined free energy in the NS limit, 
\be
\label{NS-j}
F^{\rm inst}_{\rm NS}({\bf Q}, \hbar) =\sum_{j_L, j_R \ge 0} \sum_{w\ge 1} \sum_{{\bf d}} 
{1\over w^2}N^{\bf d}_{j_L, j_R} {  \chi_{j_L} (q^{w/2}) \chi_{j_R} (q^{w/2}) 
\over q^{w/2} -q^{-w/2} } {\bf Q}^{w{\bf d}}.
\ee
to its perturbative expansion, 
\be
\label{ns-expansion}
F_{\rm NS}({\bf Q}, \hbar)=\sum_{n \ge 0} \hbar^{2n-1} F_n^{\rm NS} ({\bf Q}),
\ee
and we deduce that the finite part of $\widetilde{J}_c(\mu_{\rm eff})$ as $k \rightarrow 2$ is
\be
\label{jc-ns}
{1\over 8} \left( \partial_{t_1} -\partial_{t_2}\right)^2 F^{\rm inst}_0(t_1, t_2)\big|_{t_1=t_2=t} + F^{\text{NS, inst}}_1 (t).  
\ee
This involves the instanton part of the genus zero free energy of the off-diagonal local $\IP^1 \times \IP^1$, $F_0(t_1, t_2)$, and we have used the 
fact that the $n=0$ term in (\ref{ns-expansion}) is equal to the standard genus zero free energy: $F^{\rm NS}_0(t_1, t_2)= F_0(t_1, t_2)$. 
The expression (\ref{jc-ns}) also involves the instanton part of the NS genus one free energy, $F^{\rm NS}_1(t)$. 
This genus one free energy appears in the refinement of the topological string and it is different from the standard 
genus one free energy $F_1(t)$. Note that, again, only the $n=0,1$ free energies of (\ref{ns-expansion}) appear in this formula, and higher corrections are absent. 
Since the perturbative part of the grand potential is given by, 
\be
J^{(\rm p)}(\mu_{\rm eff},2)= {\mu_{\rm eff}^3 \over 3 \pi^2} + {\mu_{\rm eff} \over 4} + A(2), 
\ee
where \cite{hondao}
 \be
A(2)=  -{\zeta(3) \over 2 \pi^2}, 
 \ee
we conclude that
\be
\label{full-formula}
\ba
J(\mu, k=2)&=J^{(\rm p)}(\mu_{\rm eff},2)+ {\mu_{\rm eff}^2 \over 2 \pi^2} \partial^2_t F^{\rm inst}_0(t)
 - {\mu_{\rm eff} \over 2 \pi^2} \partial_t  F^{\rm inst}_0(t) +{1\over 4 \pi^2} F^{\rm inst}_0(t) + F^{\rm inst}_1(t) \\
&+{1\over 8} \left( \partial_{t_1} -\partial_{t_2}\right)^2 F^{\rm inst}_0(t_1, t_2)\big|_{t_1=t_2=t} + F^{\text {NS, inst}}_1 (t), 
\ea
\ee
and we have to set $t=2 \mu_{\rm eff}$, as required in (\ref{dictio2}). This expression can be further simplified. The first term in the second line seems to involve 
the off-diagonal theory, but in the Appendix A we show, by using special geometry, 
that it can be fully evaluated in the diagonal theory, and one finds, 
\be
\label{off-no}
\left( \partial_{t_1} -\partial_{t_2}\right)^2 F^{\rm inst}_0(t_1, t_2) \big|_{t_1=t_2=t}= \widetilde \varpi_1 (z), 
\ee
where $\widetilde \varpi_1 (z)$ is given in (\ref{tildo}). The resulting expression for $J(\mu, k=2)$ is completely explicit and 
it can be computed easily by using known results for local $\IP^1 \times \IP^1$. The genus 
zero free energy is well known from mirror symmetry, and we summarize its calculation in Appendix A. Also, the genus one free energies 
appearing in (\ref{full-formula}) are known in closed form. For the standard genus one topological string free energy, we have 
\be
F_1(t)= -{1\over 12} \log \left[ 64 z (1+16 z) \right] -{1\over 2
} \log \left( {K(-16z) \over \pi} \right), 
\ee
where $K(k^2)$ is the elliptic integral of the first kind (we use a notation in which the argument of the elliptic integrals is the square modulus $k^2$.) 
The instanton part is obtained after subtracting the linear 
term in $t$, 
\be
\label{f1-i}
F_1^{\rm inst}(t)= F_1(t)-{t\over 12} . 
\ee
Similarly, we have \cite{hk}
\be
F^{\rm NS}_1 (t)= {1\over 12} \log z -{1\over 24} \log(1+16 z). 
\ee
and
\be
\label{f1ns-i}
F^{\text{NS,  inst}}_1(t)=F^{\rm NS}_1 (t)+{t\over 12}. 
\ee
Note that, as compared to the standard formulae appearing in the literature, we have changed the sign of $z$ in
 agreement with our conventions above. By taking into account (\ref{off-no}), (\ref{f1-i}) and (\ref{f1ns-i}), we find
\be
\label{j-inter}
J(\mu, 2)=  A(2)+{\mu_{\rm eff}^3 \over 3 \pi^2} + {\mu \over 4}+
{\mu_{\rm eff}^2 \over 2 \pi^2} \partial^2_t F^{\rm inst}_0(t) - {\mu_{\rm eff} \over 2 \pi^2} \partial_t  F^{\rm inst}_0(t) +{1\over 4 \pi^2} F^{\rm inst}_0(t) + F_1(t) + F^{\rm NS}_1 (t). 
\ee
We can write down an even simpler formula for the modified grand potential, by using the natural quantities appearing in mirror symmetry, 
i.e. by using the full genus zero free energy of diagonal, local $\IP^1 \times \IP^1$:
\be
\label{full-f0}
F_0(t)= {t^3 \over 6} + F_0^{\rm inst}(t). 
\ee
In terms of this quantity, we find
\be
\label{final-k2}
J(\mu, 2)= A(2) +{1\over 4 \pi^2} \left( F_0 - t \partial_t F_0 +{1\over 2} t^2 \partial_t^2 F_0\right) + {\mu \over 4} + F_1(t) + F_1^{\rm NS} (t). 
\ee
Here, $F_0(t)$, $F_1(t)$ and $F_1^{\rm NS}(t)$ are the standard free energies appearing in the theory of diagonal, local $\IP^1 \times \IP^1$, and we identify 
\be
t= 2 \mu + 4 \re^{-2 \mu} \, _4F_3\left(1,1,\frac{3}{2},\frac{3}{2};2,2,2;-16 \re^{-2 \mu} \right), 
\ee
as prescribed by (\ref{dictio2}) and (\ref{muk2}). Using (\ref{final-k2}), it is easy to calculate $J(\mu, 2)$ in an expansion at large $\mu$ 
(in the Calabi--Yau context, this is just the large radius expansion), and check that it reproduces 
the results for the modified grand potential presented in \cite{hmo,hmo2}. 

Another useful formula, which only involves $\mu$, is obtained by taking a derivative of $J(\mu, 2)$ w.r.t. $\mu$. Using the following 
result for the Yukawa coupling, which can be found in for example \cite{dmp}, 
\be
 \partial^3_t F^{\rm inst}_0(t) = -1+\frac{\pi ^3}{8 (16 z+1) K(-16 z)^3},
 \ee
one obtains 
 \be
 {\partial J (\mu, 2)\over \partial \mu}= {1\over 4} +{1\over 1+1 6 z} \left[  {1 \over 4} \left( {\mu_{\rm eff} \over K(-16 z)} \right)^2  +{1\over 2} {E(-16 z) \over K(-16 z)} \right] -{1\over 2} {1+ 8 z \over 1+ 16 z}. 
 \ee
 Here, $E(k^2)$ is the elliptic integral of the second kind, and we remind that $z$ is related to $\mu$ through (\ref{dictioz}). This expression is very convenient in order to 
 obtain an explicit expansion at large $\mu$.

\subsection{ABJM with $k=1$}

The calculation for $k=1$ is very similar to what we have done, but there are also some important differences. First of all, the effective 
chemical potential is now given by \cite{hmo3}
\be
 \mu_{\rm eff}=\mu + \re^{-4 \mu } \, _4F_3\left(1,1,\frac{3}{2},\frac{3}{2};2,2,2;-16 \re^{-4 \mu }\right), 
 \ee
so in order to use the special geometry of local $\IP^1 \times \IP^1$ we should identify
\be
\label{tzk1}
  t= 4 \mu_{\rm eff}, \qquad  z= \re^{-4\mu}. 
  \ee
The perturbative part of the grand potential is now given by 
\be
 J^{(\rm p)}(\mu,1)= { 2 \mu^3 \over 3 \pi^2} + 3{\mu \over 8} + A(1), 
 \ee
 where $A(1)$ is known in closed form \cite{ho},
 \be
 A(1)= \frac{\log (2)}{4}-\frac{\zeta (3)}{8 \pi ^2}. 
 \ee
We can now calculate the non-perturbative part, as we did for $k=2$. As before, the worldsheet instanton contributions with $g\ge2$ vanish. One finds, 
\be\ba
J(\mu, k=1)&={ 2 \mu_{\rm eff}^3 \over 3 \pi^2} + {3 \mu_{\rm eff} \over 8} + A(1)+ {\mu_{\rm eff}^2 \over 2 \pi^2} \partial^2_t F_0^{\rm inst} (t) - {\mu_{\rm eff} \over 4 \pi^2} \partial_t  F_0^{\rm inst} (t) +{1\over 16 \pi^2} F_0^{\rm inst} (t) + F_1^{\rm inst} (t)\\
&+{1\over 32} \left( \partial_{t_1} -\partial_{t_2}\right)^2 F_0^{\rm inst} (t_1, t_2)\big|_{t_1=t_2=t} + {1\over 4}F_{(1,0)}^{\rm inst} (t)+ f\left(\mu_{\rm eff}\right), 
\ea\ee
where we use the identification (\ref{tzk1}) and we find an additional term which was not present in the case $k=2$, 
 \be 
 \ba
&  f\left(\mu_{\rm eff}\right)\\
&=\sum\limits_{w,d_1,d_2} \sum\limits_{j_L, j_R} \left((-1)^{m_L} (m_L-m_R)+m_L+m_R\right) \ri^{d_1-d_2+m_L+m_R+1} N^{d_1,d_2}_{j_L,j_R} { \re^{-2(d_1+d_2)(2w+1)\mu_{\rm eff}} 
  \over 16(2w+1)}.
  \ea
  \ee
It turns out that this function has a simple expression in terms of genus zero and genus one refined string amplitudes, 
\be
 f\left(\mu_{\rm eff}\right)={1\over 16} \left( \partial_{t_1} -\partial_{t_2}\right)^2 F_0^{\rm inst} (t_1, t_2)\big|_{t_1=t_2=t} +{3\over 4}F_{1}^{\text{NS, inst}} (t)=-{1\over 32}\log (1+16 z). 
\ee
We have verified this identity in various ways, but we do not have a proof of it. It would be interesting to fill 
out this loophole in our calculations. Using this identity, we find the following 
closed formula for the modified grand potential 
\be 
\label{fullj1} 
\ba J(\mu, k=1)&= \frac{2 \mu_{\rm eff} ^3}{3 \pi ^2}+{3\mu\over 8}+A(1)+ {\mu_{\rm eff}^2 \over 2 \pi^2} \partial^2_t F^{\rm inst}_0(t) - {\mu_{\rm eff} \over 4 \pi^2} \partial_t  F^{\rm inst}_0(t) +{1\over 16 \pi^2} F^{\rm inst}_0(t)\\
&  + F_1(t) + F^{\rm NS}_1(t).
 \ea\ee
Like in the case of $k=2$, this can be written in a slightly more compact form by using the full genus zero free energy (\ref{full-f0}):
\be 
J(\mu, 1)=A(1)+ {1\over 16 \pi^2}\left(F_0-t \partial_t F_0+{1\over 2}t^2 \partial^2_t F_0\right)+{3\mu\over 8} +F_1(t)+F_1^{\rm NS}(t).
\ee
Of course, here we have to relate $t$ to $\mu$ through (\ref{tzk1}). The derivative of $J(\mu,1)$ w.r.t. $\mu$ has a simple expression, 
\be 
\label{dj1} {\partial J(\mu, 1)  \over \partial \mu}={3\over 8} +{1\over 1+16 z} \left[ {1\over 2} \left( { \mu_{\rm eff}  \over K(-16 z)}\right)^2+ { E(-16 z) \over K(-16 z)} \right]- {1+8z \over 1+16 z}. 
\ee
We have verified explicitly that the large $\mu$ expansion of the modified grand potential given by these formulae agrees with the results in \cite{hmo2,hmo3}. 

We have then seen that the grand potential of ABJM theory simplifies enormously for $k=1, 2$. 
From the point of view of the topological string, we have a sort of  ``non-renormalization" result for these values of $k$, 
since only tree-level ($g=n=0$) and one-loop ($g=n=1$) free energies contribute 
to the final result. It is easy to see that, in ABJM theory with $k\ge 3$, all the higher genus amplitudes 
will contribute to $J(\mu, k)$. Therefore, only when $k=1,2$ do we find a simplification. It is quite 
reassuring that these are precisely the values that lead to extended $\CN=8$ supersymmetry. 

\subsection{ABJ with $M=1$ and $k=2$}

The calculation of the grand potential in this case is again very similar to the previous ones. 
For $M=1$ and even $k$, the effective chemical potential is given by \cite{hondao}
\be
 \mu_{\rm eff}=\mu-2 \re^{-2 \mu } \, _4F_3\left(1,1,\frac{3}{2},\frac{3}{2};2,2,2;16 \re^{-2 \mu } \right).
 \ee
 %
This means that the dictionary with special geometry takes now the form 
\be\label{tm} t=2 \mu_{\rm eff}+\ri \pi, \qquad  z=-\re^{-2\mu}. 
\ee
The computation of the modified grand potential is very similar to the case of ABJM with $k=2$, and one finds:
\be \label{jnf2}
J(\mu, 2;1)=  A(2,1)+{\mu_{\rm eff}^3 \over 3 \pi^2} +  
{\mu_{\rm eff}^2 \over 2 \pi^2} \partial^2_t F^{\rm inst}_0(t) - {\mu_{\rm eff} \over 2 \pi^2} \partial_t  F^{\rm inst}_0(t) +{1\over 4 \pi^2} F^{\rm inst}_0(t) + F_1(t) + F^{\rm NS}_1 (t), 
\ee
where 
\be
A(2,1)=\frac{1}{2} \left(\log (2)-\frac{\zeta (3)}{ \pi ^2}\right). 
\ee
Let us now define the genus zero free energy by 
\be F_0(t)={(t-\ri \pi)^3\over 6}+F_0^{\rm inst}(t).
\ee 
Then, the result (\ref{jnf2}) can be written in a more compact form as
\be 
\label{jnf2lr}
J(\mu, 2;1)= A(2,1)+{1\over 4 \pi^2}\left( F_0-(t-\ri \pi)\partial_t F_0+{1\over 2}(t-\ri \pi)^2\partial_t^2 F_0\right)+F_1(t) +F_1^{\rm NS}(t). 
\ee
Like in the previous cases, the derivative of this function can be written in closed form in terms of elliptic integrals and $\mu_{\rm eff}$, 
\be 
\label{dj2}
{\partial J(\mu, 2;1) \over \partial \mu} ={1\over 1+16 z} \left[ {1\over 4} \left( { \mu_{\rm eff}  \over K(-16 z)}\right)^2+ {1\over 2} { E(-16 z) \over K(-16 z)} \right]- {1\over 2} {1+8z \over 1+16 z}.
\ee
We have verified that these expressions for the modified grand potential, once expanded at large $\mu$, agree with the results obtained in \cite{hondao}.

\sectiono{Generating functionals for the partition function}

In the previous section we have written down closed formulae for the modified grand potential of $\CN=8$ ABJ(M) theories. Would it be possible to obtain 
from these results explicit generating functionals for the partition functions? In other words, can we compute the standard grand canonical partition function of these theories? 
In this section we will show that indeed this is possible, and that the answer is related in an interesting way to the non-perturbative 
partition functions studied in \cite{bde,eynard, em}. 

\subsection{ABJM with $k=2$} 

As in the previous section, let us start with the simplest case, namely ABJM theory with $k=2$. In order to evaluate $\Xi(\mu, k=2)$, we should use the formula (\ref{naif-gp}), together with the 
explicit expressions (\ref{j-inter}), (\ref{final-k2}) for the modified grand potential. Notice that the instanton part of (\ref{j-inter}) depends on $\mu$ through $z$, so it is left invariant by the shift 
\be
\label{mus}
\mu \rightarrow \mu+ 2 \pi \ri n, \qquad n \in \IZ,
\ee
Therefore, the shift only affects $\mu_{\rm eff}$. An easy calculation shows that
\be
\label{exp-shift}
\exp\left[ J(\mu+ 2 \pi \ri n, 2) \right]= \exp \left(J(\mu, 2) \right)\exp \left[ \pi \ri n^2 \tau +  2 \pi \ri n \left( \xi -{1\over 12}\right) \right], 
\ee
where 
\be\label{tau-f}
\tau= {2 \ri \over  \pi} \partial_t^2 F _0 =-{\ri \over \pi} {\varpi_2'(z) \over \varpi'_1(z)}
\ee
and
\be
\label{our-xi}
\xi= {1\over 2 \pi^2} \left(t \partial_t^2 F_0 - \partial_t F_0 \right).
\ee
In (\ref{tau-f}), $\varpi_{1,2}(z)$ are the periods of local $\IP^1 \times \IP^1$, and they are defined in (\ref{vares}). 
Notice that in calculating $J(\mu+ 2 \pi \ri n, 2)$ one obtains a cubic term in $n^3$, but in deriving (\ref{exp-shift}) we used that
\be
\label{cubic-s}
\exp\left( - {8 \pi n^3 \ri \over 3} \right)= \exp\left( - {2 \pi n \ri \over 3} \right), \qquad n \in \IZ. 
\ee

We now recognize the form of the second factor in (\ref{exp-shift}): it is the standard summand of a Jacobi theta function. Of course, in order for this interpretation to 
be correct, one needs ${\rm Im}(\tau)>0$. But the $\tau$ appearing here is (up to an overall factor of $2$ and an integer shift) the modular parameter of the spectral curve describing the 
planar solution of ABJM theory \cite{dmp}. Therefore, the resulting theta function is well-defined, and we finally obtain:
\be
\label{xi-ex}
\Xi (\mu, k=2)= \exp\left(J(\mu,2)\right)\vartheta_3 \left(\xi-{1\over 12}, \tau\right),
 \ee
 where $\vartheta_3(v, \tau)$ is the Jacobi theta function, defined in (\ref{jtheta}).

 The function (\ref{xi-ex}) is very similar to the ``non-perturbative partition function" $Z_{\alpha, \beta}(\Sigma)$  
 introduced in \cite{bde,eynard} and further studied in \cite{em}. Let us briefly review its construction, following 
 the notations of \cite{em} (see also \cite{mmlargen} for an overview in the context of matrix model asymptotics). 
 The function $Z_{\alpha, \beta}(\Sigma)$ is canonically associated to 
 a spectral curve $\Sigma$, together with a choice of meromorphic differential 
 \be
 \label{ydx}
 \lambda =y(x) \rd x.
 \ee
 The basic ingredients in constructing this function are the free energies ${\bf F}_g$, determined by the pair $\left( \Sigma, \lambda \right)$ via special geometry and 
 the topological recursion of \cite{eo} (we use a boldface notation since these free energies differ from the ones used above in overall 
 normalizations). Let us focus on the case in which $\Sigma$ has genus one, which is the relevant one for us. 
 Given two symplectically conjugated cycles on $\Sigma$, $\CA$, $\CB$, one defines the genus zero free energy ${\bf F}_0(\epsilon)$ from the standard relationships in 
 special geometry, 
 \be
\label{eps}
\epsilon = {1\over 2\pi \ri}\,\oint_{\CA} \lambda, \qquad \textbf{F}'_0  = \oint_{\CB} \lambda, 
\ee
where the $'$ denotes a derivative w.r.t. $\epsilon$. To construct the ``non-perturbative partition function," one needs in addition the theta function with characteristics $\alpha$, $\beta$, 
\be
\vartheta\left[^\alpha_\beta\right](\xi|\tau)= \sum_{n \in \IZ }
\exp\left[ \ri \pi (n+\alpha) \tau (n+\alpha) + 2\pi \ri (n+\alpha)(\xi+\beta)\right], 
\ee
as well as a modified theta function which depends on an additional parameter $N$, 
\be
\label{Thetatheta}
\Theta_{\alpha,\beta}=\exp\left[ -N^2\left(  \epsilon \textbf{F}_0'-\pi \ri  \epsilon^2 \tau \right)\right] 
\,\, \vartheta\left[^\alpha_\beta\right](\xi|\tau).
\ee
The argument of the theta function is given by 
\be
\label{xivalue}
\xi=N \left( {\textbf{F}'_0 \over 2\pi \ri}  - \tau \epsilon \right).
\ee
We recall that, under a modular transformation
\be
\Gamma=\begin{pmatrix}a & b\\ c & d \end{pmatrix} \in {\rm SL}(2,\IZ), 
\ee
where $a,b,c,d \in \IZ$ and $ad-bc=1$, we have 
\be
\tau \rightarrow \bar \tau ={ a \tau+b \over c\tau+d}, 
\ee
and
\be
\label{xi-trans}
\xi \rightarrow \bar \xi = {\xi \over c \tau+d},
\ee
therefore $\xi$ is a modular form of weight $-1$, as required from the argument of a theta function.

The non-perturbative partition function $Z_{\alpha, \beta}(\Sigma)$ of \cite{bde,eynard,em} is defined by a formal $1/N$ expansion, 
and its leading term, which we will denote as $\CT_{\alpha, \beta}(\Sigma)$, is given by  
 \be
 \label{leading-z}
\CT_{\alpha, \beta}(\Sigma)= \exp\left( N^2 \textbf{F}_0 + \textbf{F}_1\right) \Theta_{\alpha, \beta}. 
 \ee
The full function $Z_{\alpha, \beta}(\Sigma)$ is given by (\ref{leading-z}), plus an infinite series  of $1/N$ corrections which involve 
the higher genus free energies ${\bf F}_g$, with $g \ge 2$. These corrections will not be needed in our case. 
One of the most important properties of (\ref{leading-z}), proved in \cite{em}, is that it is essentially modular invariant: 
it transforms into itself, up to a phase and a change of characteristic of 
 the theta function. 
 
It is easy to see that the function (\ref{xi-ex}) can be put in the form of (\ref{leading-z}), up to an overall function of $z$. To see this, we take into account the following relationship between 
the quantities appearing in the special geometry of local $\IP^1 \times \IP^1$, and those used in \cite{em}:
\be
\label{t-rules}
t= -2 \epsilon, \qquad F_0 = -{\bf F}_0, \qquad F_1 = {\bf F}_1. 
\ee
In particular, we have that $\partial_tF_0= {\bf F}'_0/2$. The modular properties of $t$ and $\partial_t F_0$ can 
be obtained from those written down in \cite{em} by using (\ref{t-rules}), and the quantity (\ref{xivalue}) coincides with (\ref{our-xi}). 
Once (\ref{t-rules}) is taken into account, we find that (\ref{xi-ex}) can be written as 
\be
\label{ximu-gen}
\Xi(\mu,2)= \exp \left( {\mu \over 4} + F_1^{\rm NS} \right) \CT_{0, \beta}(\Sigma), 
\ee
with 
\be
\beta=-{1\over 12}, \qquad N={1\over 2 \pi \ri}. 
\ee
We now note that the prefactor 
\be
\label{prefactor}
 \exp \left( {\mu \over 4} +  F_1^{\rm NS} \right)
 \ee
only depends on the ``bare" coordinate $z$, therefore it is modular invariant. We conclude that $\Xi(\mu, 2)$ has the same properties under a modular transformation than 
the ``non-perturbative partition function" or its leading order term (\ref{leading-z}): it is modular invariant, up to a phase and a change of characteristic. 
It is interesting to note that, by the general theory of Fredholm determinants, (\ref{std2}) should be an entire function of $\kappa$, which is not manifest from the explicit expression 
(\ref{ximu-gen}) (at least, it is not manifest to us.) 

We should also note that, in our case, there is an additional subtlety w.r.t. to the analysis of \cite{em}. The reason is that 
in local mirror symmetry there are three different periods, and not two. One of the 
periods is constant and usually it does not play a crucial r\^ole, but its presence means that the non-trivial 
periods can be shifted by constants when making a modular transformation. One consequence 
of the presence of such shifts is that the quantity which transforms as (\ref{xi-trans}) is not $\xi$, but rather the shifted quantity $\xi-1/12$ appearing in the argument of the theta function. Therefore, 
the natural theta function is the one written down in (\ref{xi-ex}), with characteristics $\alpha=\beta=0$. 

The relation to the ``non-perturbative perturbative function" of \cite{bde,eynard,em} is not that surprising, 
since this function is obtained by summing the matrix model partition function over all possible filling fractions, and the resulting sum is 
very similar to the sum over $n$ appearing in (\ref{naif-gp}). There are however two crucial differences with the formalism of \cite{bde,eynard,em}. First, in the sum 
in (\ref{naif-gp}) there is a factor of $\ri$ in front of $n$, which means that one is considering imaginary shifts of the modulus, while in the ``non-perturbative partition 
function" one sums over real shifts of the modulus. As a consequence, the theta function has an oscillatory behavior in the large radius region, while in the 
formalism of \cite{bde,eynard,em} one has to require in addition a Boutroux condition on the moduli. 
The other important difference is that, in computing the grand canonical partition function, we must 
include the contribution of membrane instantons, which appear in the next-to-leading NS free energy $F_1^{\rm NS}$ of (\ref{prefactor}). 
These are invisible in the 't Hooft expansion of the matrix model, 
which was the only ingredient considered in \cite{bde,eynard,em}. 
Therefore, the ``non-perturbative partition function" of these papers does {\it not} include the full non-perturbative information required in this problem.   

The expression (\ref{xi-ex}), when expanded in the fugacity (\ref{fugacity}) around $\kappa=0$, should give the generating functional for all the partition functions $Z(N,k=2)$. 
However, the expression (\ref{xi-ex}) is not immediately useful for this purpose. The reason is that the quantities 
involved in this equation are appropriate for the large radius regime, which corresponds to large and positive chemical 
potential, 
\be
\mu \rightarrow \infty, 
\ee
while the expansion around $\kappa=0$ is an expansion around 
\be
\mu \rightarrow -\infty. 
\ee
In the language of special geometry, the expansion around $\kappa=0$ 
is an expansion around the orbifold point $z \rightarrow  \infty$, while the expansion considered in the previous section was an expansion 
at large radius $z\rightarrow 0$. It is known that these two points are related by a modular transformation \cite{abk}, and this fact was heavily exploited in \cite{dmp} to 
determine the full $1/N$ expansion of the free energy. Since the relevant modular transformation is essentially an $S$-transformation, 
it is convenient to introduce the orbifold modular parameter, 
\be
\bar \tau= -{1\over \tau}= {\ri \over 2}  {K'\left(-\kappa^2/16\right)\over K \left(-\kappa^2/16\right)} -{1\over 2}. 
\ee
We also introduce the $S$-transform of $\xi-1/12$, 
\be
\bar \xi = {\xi -{1\over 12} \over \tau}.
\ee
The almost modular invariance of $\Xi(\mu,2)$ says that it should be possible to write it in the original form (\ref{xi-ex}), (\ref{final-k2}), but in terms of quantities appropriated to 
the orbifold frame. In particular, the function 
\be
F_0 - t \partial_t F_0 +{1\over 2} t^2 \partial_t^2 F_0
\ee
appearing in (\ref{final-k2}), should be written in the orbifold frame.
A convenient basis of periods around the orbifold point is the one featuring in the planar solution of ABJM theory worked out in \cite{dmp}, 
\be 
\label{standard}
\ba
\lambda&={\kappa \over 8 \pi}  {~}_3F_2\left(\frac{1}{2},\frac{1}{2},\frac{1}{2};1,\frac{3}{2};-\frac{\kappa^2
   }{16}\right), 
 \\
 \partial_\lambda \CF_0 (\lambda)&= { \kappa \over 4 } G^{2,3}_{3,3} \left( \begin{array}{ccc} {1\over 2}, & {1\over 2},& {1\over 2} \\ 0, & 0,&-{1\over 2} \end{array} \biggl| -{\kappa^2\over 16}\right)+ { \pi^2 \ri \kappa\over 2}   {~}_3F_2\left(\frac{1}{2},\frac{1}{2},\frac{1}{2};1,\frac{3}{2};-\frac{\kappa^2
   }{16}\right),  \ea
\ee
where we have denoted by $\CF_0(\lambda)$ the orbifold frame genus zero free energy of \cite{dmp}. Its expansion around $\lambda=0$ is given by 
\be
\CF_0(\lambda)=-4 \pi ^2 \lambda ^2 \left(\log (2 \pi  \lambda )-\frac{3}{2}-\log (4)\right)+\cdots
\ee
We also have that, 
\be
\partial_\lambda^2 \CF_0 (\lambda)= - 8 \pi^3 \ri \bar \tau. 
\ee
Note that the periods in (\ref{standard}) are analytic continuations of the periods at large radius, see (\ref{per-anc}) for the precise relationship. This was 
useful in the calculations of \cite{dmp} to relate the strong coupling and the weak coupling expansions 
of the planar free energy. We will now use this analytic continuation, together with the modular 
transformation, to relate the quantities appearing in (\ref{xi-ex}) to orbifold quantities. Using the modular properties of $\xi$, 
we should expect that $\bar \xi$ is given by an expression similar to (\ref{our-xi}), but in the orbifold frame. Indeed, a simple computation shows that
\be
\bar \xi= {\ri \over 4 \pi^3} \left( \lambda \partial_\lambda^2 \CF_0 (\lambda) - \partial_\lambda \CF_0 (\lambda)\right), 
\ee
and we find, for (\ref{ximu-gen}), 
\be 
\label{std2}
\Xi(\mu, 2)=  \exp \left\{ {\mu \over 4}+\CF_1+F_1^{\rm NS} -{1\over  \pi^2}  \left( \CF_0(\lambda)-\lambda \partial_{\lambda}\CF_0(\lambda)+{\lambda^2 \over 2} 
\partial^2_{\lambda}\CF_0(\lambda)\right) \right\}
\vartheta_3(\bar \xi, \bar \tau). 
\ee
Here, $\CF_1$ is the genus one free energy at the orbifold point, which is given by \cite{abk,dmp}
\be
\label{F1st} \CF_1=- \log \eta\left( 2 \bar \tau\right)-{1\over 2}\log 2. 
\ee
The linear term $\mu/4$ appearing 
in (\ref{std2}) cancels against similar terms in $\CF_1$ and $F_1^{\rm NS}$, and all the other quantities appearing here have power series expansions 
in $\kappa$, around $\kappa=0$. For example, the theta function has the following expansion, 
\be
\vartheta_3(\bar \xi, \bar \tau)=1+\frac{\kappa }{8}+\left(\frac{1}{64 \pi ^4}-\frac{1}{512}\right) \kappa ^3+ \frac{\kappa ^4}{32768}+\mathcal{O}(\kappa^5),
\ee
and
\be
\ba
-{1\over \pi^2} \left( \CF_0(\lambda)-\lambda \partial_{\lambda}\CF_0(\lambda)
+{\lambda^2 \over 2} \partial^2_{\lambda}\CF_0(\lambda)\right)& =
 \frac{\kappa ^2}{32 \pi ^2} -\frac{\kappa ^4}{1536 \pi ^2}+\frac{199 \kappa ^6}{8847360 \pi ^2}+\mathcal{O}(\kappa^8),\\
 {\mu \over 4}+\CF_1+F_1^{\rm NS} &=\frac{\kappa ^4}{32768}-\frac{\kappa ^6}{524288}+\mathcal{O}(\kappa^8). \\
 \ea
 \ee
We can now put all these ingredients together and compute the expansion of $\Xi(\mu,2)$, which gives then the values of $Z(N,2)$ for $N=1,2, \cdots$. We find, 
\be
\Xi(\mu,2)= 1+\frac{\kappa }{8}+\frac{\kappa ^2}{32 \pi ^2}+\frac{\left(10-\pi ^2\right) \kappa^3}{512 \pi ^2} 
 +\frac{\left(24-32 \pi ^2+3 \pi ^4\right) \kappa^4}{49152 \pi ^4}+\CO\left(\kappa ^5\right).  
   \ee
From this expansion we can read, 
\be
Z(1, 2)={1\over 8}, \qquad Z(2,2)= {1\over 32 \pi^2}, \qquad Z(3,2)=\frac{10-\pi ^2}{512 \pi ^2}, 
\ee
and so on. This agrees with the results obtained in \cite{hmo2}. Note that the coefficients in this power series are polynomials in $1/\pi$ with 
rational coefficients, as noted in \cite{py,hmo}. Our formula (\ref{std2}) makes this essentially manifest. Of course, we can push the calculation up to any 
order in $\kappa$, and we have checked that we reproduce all known values of $Z(N,2)$. 
   
\subsection{ABJM with $k=1$ }
For $k=1$, the relevant dictionary is (\ref{tzk1}). The bare modulus $z$ is then related to the fugacity as
\be
\label{z4} z= {1\over \kappa^4}.
\ee
A computation similar to the one for $k=2$ gives
\be 
\label{xi1}
\Xi (\mu, 1)=  \exp \left( J(\mu, 1) \right) \vartheta_3 \left( {\xi \over2}-{7\over 24}, \tau \right). 
\ee
Like before, we would like to obtain from this expression a generating functional of the partition functions $Z(N,1)$. We use again modular invariance and re-express the above result in terms of 
orbifold quantities. After using some theta functions identities, listed in (\ref{theta-ids}), one finds, 
\be 
\label{std1} 
\ba
\Xi (\mu, 1)=& \exp\left[ {3 \mu\over 8}-{3\over 4}\log 2 +\CF_1 +F_1^{\rm NS}
 - {1\over 4 \pi^2}  \left(  \CF_0(\lambda)- \lambda \partial_{\lambda}\CF_0(\lambda)+{\lambda^2\over 2}  \partial^2_{\lambda}\CF_0(\lambda)\right) \right] \\
& \times \left( \vartheta_2\left( \bar \xi/4, \bar \tau/4 \right) +\ri \vartheta_1\left(\bar \xi/4, \bar \tau/4 \right) \right).
\ea
\ee
All the quantities appearing here are the same ones appearing in the formula (\ref{std2}) for $k=2$, with the only difference that we have to change 
\be
\kappa \rightarrow \kappa^2, 
\ee
due to the relation (\ref{z4}). We have the following expansions around $\kappa=0$,
\be
\ba
  \label{thex}
 {3 \mu\over 8}-{3\over 4}\log 2+F_1^{\rm NS} +\CF_1 &=-\frac{\log (\kappa )}{8} - \frac{3 \log (2)}{4}+ \frac{\kappa ^8}{32768}+\mathcal{O}(\kappa^{12}), \\
 \vartheta_2\left( \bar \xi/4, \bar \tau/4 \right) +\ri \vartheta_1\left(\bar \xi/4, \bar \tau/4 \right) &= \exp\left( \frac{\log (\kappa )}{8} + \frac{3 \log (2)}{4} \right) \left(1+\frac{\kappa }{4}+ \frac{\kappa ^2}{16 \pi} + \mathcal{O}(\kappa^3) \right),
 \ea
 \ee
 and we see that, as in the case of $k=2$, all the $\log(\kappa)$ terms cancel (as well as the constant terms.) Note that in this calculation the 
 dictionary relating $z$ to $\kappa$ is given by (\ref{z4}). Putting everything together, we find,
 \be
 \Xi(\mu, 1)= 1+\frac{\kappa }{4}+\frac{\kappa ^2}{16 \pi }+\frac{(\pi -3) \kappa ^3}{64 \pi }+{ (10-\pi^2) \kappa^4 \over 1024 \pi^2}+\CO\left(\kappa ^5\right), 
   \ee
   which reproduces the known results for $Z(N,1)$ in \cite{py,hmo,hmo2}, for the very first $N$ (see (\ref{zn1})). Again, 
   we can push the computation up to arbitrary order and reproduce all known values of $Z(N,1)$ with the above 
   generating function. 

\subsection{ABJ with $M=1$ and $k=2$ }
In this case, the modulus $z$ is related to the fugacity by
\be
\label{zabj} 
z= -{1\over \kappa^2}. 
\ee
As in the previous cases, we find the following explicit formula, 
\be
 \label{jf} 
 \Xi (\mu, 2;1)=\exp\left( J(\mu, 2;1)\right)  \vartheta_{3}\left (\xi-{\tau \over 4} -{7\over 12}, \tau \right). 
 \ee
 We can now use modular invariance and standard transformations of the theta functions to write this quantity in the orbifold frame,
 \be 
 \label{stdnf}
 \ba
 \Xi(\mu, 2;1)&=\exp \left[ {\log 2 \over 2}+\CF_1+F_1^{\rm NS}
  -{1\over  \pi^2}  \left( \CF_0(\lambda)-\lambda \partial_{\lambda}\CF_0(\lambda)+ {\lambda^2\over 2}  \partial^2_{\lambda}\CF_0(\lambda)\right)  \right] \\
& \times \vartheta_1\left(\bar \xi+{1\over 4}, \bar \tau \right).
\ea
 \ee
The quantities that appear here are the same ones appearing in (\ref{std2}), but with the change
 \be
 \kappa \rightarrow \ri \kappa. 
 \ee
 Like before, there is a cancellation of $\log(\kappa)$ terms between the theta functions and the genus one free energies. We have the following expansions around $\kappa=0$, 
 \be
 \ba
\CF_1+F_1^{\rm NS}&=- \frac{\log (\kappa )}{4}+{  \pi  \ri \over 8} +\frac{\kappa ^4}{32768}+\mathcal{O}(\kappa^{6}), \\
 \vartheta_1\left(\bar \xi+{1\over 4},\bar \tau \right)&=\exp\left( {\log (\kappa) \over 4} -{ \pi \ri \over 8}-{\log 2 \over 2} \right) 
 \left( 1 +\frac{\kappa }{4 \pi }+ \frac{1}{128} \left(1-\frac{4}{\pi ^2}\right) \kappa ^2+ \CO\left(\kappa^3\right) \right).
 \ea \ee
 We then find, 
 \be
 \Xi(\mu, 2;1)= 1+\frac{\kappa }{4 \pi }+\frac{\left( \pi^2 -8 \right) \kappa^2}{128 \pi^2}+\frac{\left(5 \pi ^2-48\right)
   \kappa ^3}{4608 \pi ^3}+\frac{\left(480-848 \pi ^2+81 \pi ^4\right) \kappa ^4}{294912 \pi ^4}+\CO\left(\kappa
   ^5\right), 
   \ee
   which agrees with the result obtained in \cite{hondao,matsumori} for the very first $N$. We have checked that the generating function (\ref{stdnf}) 
   reproduces all known values of $Z(N,2;1)$.
   
\sectiono{Exact quantization conditions for the $\CN=8$ Fermi gas}

In this section, we will consider the spectrum of the integral operator with kernel (\ref{densitymat}), in the maximally supersymmetric 
$\CN=8$ theories. This is just the spectrum of the 
one-particle Hamiltonian of the ideal Fermi gas of \cite{mp}. It was already 
suspected in \cite{mp} that, in the maximally supersymmetric cases, this spectrum could be found in some relatively simple form, 
and it was speculated that a connection to integrable systems 
would be instrumental. We will now show that the explicit results for the grand potential obtained in the previous section lead to 
{\it exact quantization conditions} for the spectrum of the one-particle Hamiltonian. 

It follows from the product formula (\ref{ximu}) that the zeroes of the grand potential occur when the chemical potential is given by 
\be
\label{mu-zer}
\mu=E_n \pm \pi \ri, 
\ee
where $E_n$ is the $n$-th energy level. Since the grand potential was obtained by shifting $\mu$ by $2 \pi \ri n$ and summing over all possible $n$, 
evaluating it at (\ref{mu-zer}) is equivalent to evaluate it at $\mu=E$ and replacing $n$ in the sum (\ref{naif-gp}) by $n\pm 1/2$. This has the effect of 
changing the characteristic of the theta function: it transforms $\vartheta_3$ into $\vartheta_2$. We conclude that, in the case of ABJM theory with $k=2$, the 
spectral determinant (\ref{xi-ex}) becomes
\be
\Xi\left(E\pm \pi \ri , k=2\right)=\exp\left(J(\mu,2)\right)\vartheta_2 \left(\xi-{1\over 12}, \tau\right). 
\ee
When does this vanish? As it is well-known (see for example \cite{akhiezer}), the theta function appearing here is an oscillatory 
function, proportional to 
\be
\cos\left( \pi \left( \xi-{1\over 12}\right) \right). 
\ee
This vanishes when 
\be
\label{quant-general}
\xi(E)-{1\over 12}= n+{1\over 2}, \qquad n=0, 1, 2, \cdots 
\ee
This is our exact quantization condition. As noted in \cite{km}, quantization conditions for ABJM theory can be written in the quantum-corrected 
Bohr--Sommerfeld form 
\be
{\rm vol}(E; k)= 4 \pi^2  k\left( n+{1\over 2}\right), \qquad n=0,1,2, \cdots, 
\ee
where ${\rm vol}(E;k)$ is usually called the quantum volume. By using the explicit expressions for the function $\xi(E)$, one finds an explicit formula for the 
quantum volume when $k=2$, 
\be
\ba
{\rm vol}(E;2) &= 8\pi {K(1+16 \re^{-2E} ) \over K(-16  \re^{-2E})} \left( E + 2  \re^{-2E} \, _4F_3\left(1,1,\frac{3}{2},\frac{3}{2};2,2,2;-16  \re^{-2E}\right) \right)
\\ & -{4\over  \pi} 
G_{3,3}^{3,2}\left(-16  \re^{-2E}\left|
\begin{array}{c}
 \frac{1}{2},\frac{1}{2},1 \\
 0,0,0
\end{array}
\right.\right).
\ea
\ee
Using this formula, we can compute the energy levels of the Fermi gas 
with arbitrary precision. We find, for example, 
for the ground state energy, 
\be
E_0=2.3623774930139632119156...
\ee
which agrees with a numerical calculation with $20$ significant digits. In principle, the theta function has additional zeros, given by
\be
\label{quant-general-m}
\xi(E)-{1\over 12}= n+{1\over 2} + m \tau, \qquad m,n \in \IZ. 
\ee
However,  when $m\not=0$ and/or $n<0$, this equation does not have solutions in the complex $E$ plane. This is as it should be, since it follows from (\ref{ximu}) that the 
only possible zeroes of the grand canonical partition function are related to the energy levels by (\ref{mu-zer}). 

Similar considerations apply to the other ABJ(M) theories with maximal $\CN=8$ supersymmetry. For ABJM with $k=1$, one finds, by looking at at the vanishing of the 
theta function in (\ref{xi1}), the following exact quantization condition:
\be 
{\xi(E) \over 2 }-{7\over 24}=n+{1\over 2}, \qquad n=0, 1, 2, \cdots
\ee
This leads to the following explicit formula for the quantum volume, 
\be
\ba
{\rm Vol}(E, k=1)&=-\pi^2 +4 \pi  {K(1+16  \re^{-4E})\over K(-16\re^{-4E})} \left( E+ \re^{-4 E }  \, _4F_3\left(1,1,\frac{3}{2},\frac{3}{2};2,2,2;-16 \re^{-4 E } \right) \right)\\
& -{1\over \pi} G_{3,3}^{3,2}\left(-16 \re^{-4E}\left|
\begin{array}{c}
 \frac{1}{2},\frac{1}{2},1 \\
 0,0,0
\end{array}
\right.\right). 
\ea
\ee 

Finally, we consider the maximally supersymmetric ABJ theory with $M=1$, $k=2$. Again, one can read the exact quantization condition 
from the theta function appearing in (\ref{jf}). One finds, 
 \be 
 \label{qcNf}  
 \xi -{\tau \over 4}-{7\over 12}=n+{1\over 2}, \qquad n=0, 1,2, \cdots, 
 \ee
In this case the WKB quantization condition reads \cite{kallen}
\be
{\rm vol}(E, k; M)= 4 \pi^2 k \left( n+{1\over 2}\right), \qquad n=0,1,2, \cdots, 
\ee
and one finds from (\ref{qcNf}) an exact expression for the quantum volume, 
\be 
\label{VolNf} 
\ba
{\rm Vol}(E, k=2;M=1)&= 8\pi  {K(1-16 \re^{-2E})  \over K(16 \re^{-2E} )} \left(E- 2 \re^{-2E} \, _4F_3\left(1,1,\frac{3}{2},\frac{3}{2};2,2,2;16 \re^{-2E} \right) \right)\\
&  - {4\over  \pi} G_{3,3}^{3,2}\left(16 \re^{-2E} \left|
\begin{array}{c}
 \frac{1}{2},\frac{1}{2},1 \\
 0,0,0 \\
\end{array}
\right.\right). 
\ea
\ee
This leads for example to the following value for the ground state energy, 
\be 
E_0 \approx 2.8818154299262968...
\ee
This agrees with a numerical calculation of the spectrum (and with a less precise, previous calculation in \cite{kallen}).

In \cite{km}, exact expressions were conjectured for the quantum volume function ${\rm vol}(E;k)$ of the 
ABJM Fermi gas with general $k$, and in \cite{kallen} these were generalized to ABJ theory. It can be verified by a direct calculation that the explicit expressions 
derived above from the vanishing of the theta function agree with the conjectures of \cite{km,kallen} in the maximally supersymmetric cases\footnote{It is now known that 
the conjectural expressions for the quantum volume put forward in \cite{km,kallen} receive corrections for general values of $k$ \cite{hw,ghm}. Corrected
quantization conditions for general $k$ have been proposed in \cite{ghm}, based on a generalization of the theta functions analyzed in this section.}. 

As a final comment, note that the partition functions (\ref{tanh-form}) and (\ref{abj-alt}) can be also regarded as canonical partition functions of {\it classical}, interacting one-dimensional 
gases, with an interacting potential of the form $-\log(\tanh(x/2))$ \cite{gm}. In this context, the zeroes of the grand canonical partition function, which are 
determined by our exact quantization conditions, are nothing but the Lee--Yang zeros in the fugacity plane. The connection between Lee--Yang zeros and the zeroes of 
the theta function appearing in the ``non-perturbative partition function" of \cite{bde,eynard,em} was already pointed out in section 4.7 of \cite{mmlargen}. 

\sectiono{On the quantum geometry of M-theory}

\subsection{Precision tests and quantum geometry}

Let us now come back to (\ref{naif}), which we took as our definition of the modified grand potential. In this definition, 
the ABJM partition function $Z(N,k)$ (which is in principle only defined when $N$ is a positive integer) is obtained as a Laplace transform 
of $\exp( J(\mu,k))$. In this paper we have obtained closed formulae for $J(\mu, k)$ in the maximally supersymmetric case, so we can 
check with high precision that in these cases the Airy type of integral in the r.h.s. of (\ref{naif}) reproduces the known values of $Z(N,k)$ for $k=1,2$.  In this way, 
we have been able to verify that the modified grand potential conjectured in previous works \cite{mp,hmo2,hmo3, hmmo}, when specialized to the $\CN=8$ theories, leads to the right 
values of the partition function with a precision which is in some cases of one part in $10^{500}$. 
A useful tool to perform this calculation is that, as already noted in \cite{hmo2}, one can 
expand the integrand 
in (\ref{naif}) as 
\begin{equation}
\re^{J(\mu, k)} = \re^{J^{({\rm p})} (\mu, k)} \sum_{l=1}^{\infty} \re^{-\frac{4l}{k}\mu}\sum_{n=0}^{2l} a_{l,n} \mu^n.
\end{equation}
After integration, the expansion in $\mu$ can translated into derivatives with respect to $N$, and this leads to the expression
\begin{equation}
\label{z-exp}
Z(N,k)=\frac{\re^{A(k)}}{C(k)^{1/3}}
\sum_{l=1}^{\infty} \sum_{n=0}^{2l} a_{l,n} \left(-\frac{\partial}{\partial N}\right)^n \mathrm{Ai}
\left(\frac{N+\frac{4l}{k}-B(k)}{C(k)^{1/3}}\right). 
\end{equation}
This can be regarded as an instanton expansion, since the 
behaviour of the Airy function at large argument leads to an exponential 
suppression of the higher order $l$ terms. For $k=2$, 
we have for example, 
\begin{equation} 
\label{leadingco}
\ba 
Z\left(N,2\right)=\re^{-\frac{\zeta (3)}{2 \pi ^2}}&\left\{ \pi ^{2/3} \mathrm{Ai}\left(\frac{N-1/4}{\pi ^{-2/3}}\right)+\pi^{-4/3}\left(\pi ^2 (4 N+7)+1\right)\mathrm{Ai}\left(\frac{N+7/4}{\pi ^{-2/3}}\right) \right.\\
& \left.-2\pi ^{-2/3}\mathrm{Ai}'\left(\frac{N+7/4}{\pi^{-2/3}}\right)\right\}+\CO\left(\re^{-\frac{2(N+15/4)^{3/2}}{3}}\right).
\ea \end{equation}
Of course, this type of considerations can be also made for the maximally supersymmetric ABJ theory with $M=1$ and $k=2$.

\begin{figure}
\center
\includegraphics[height=6cm]{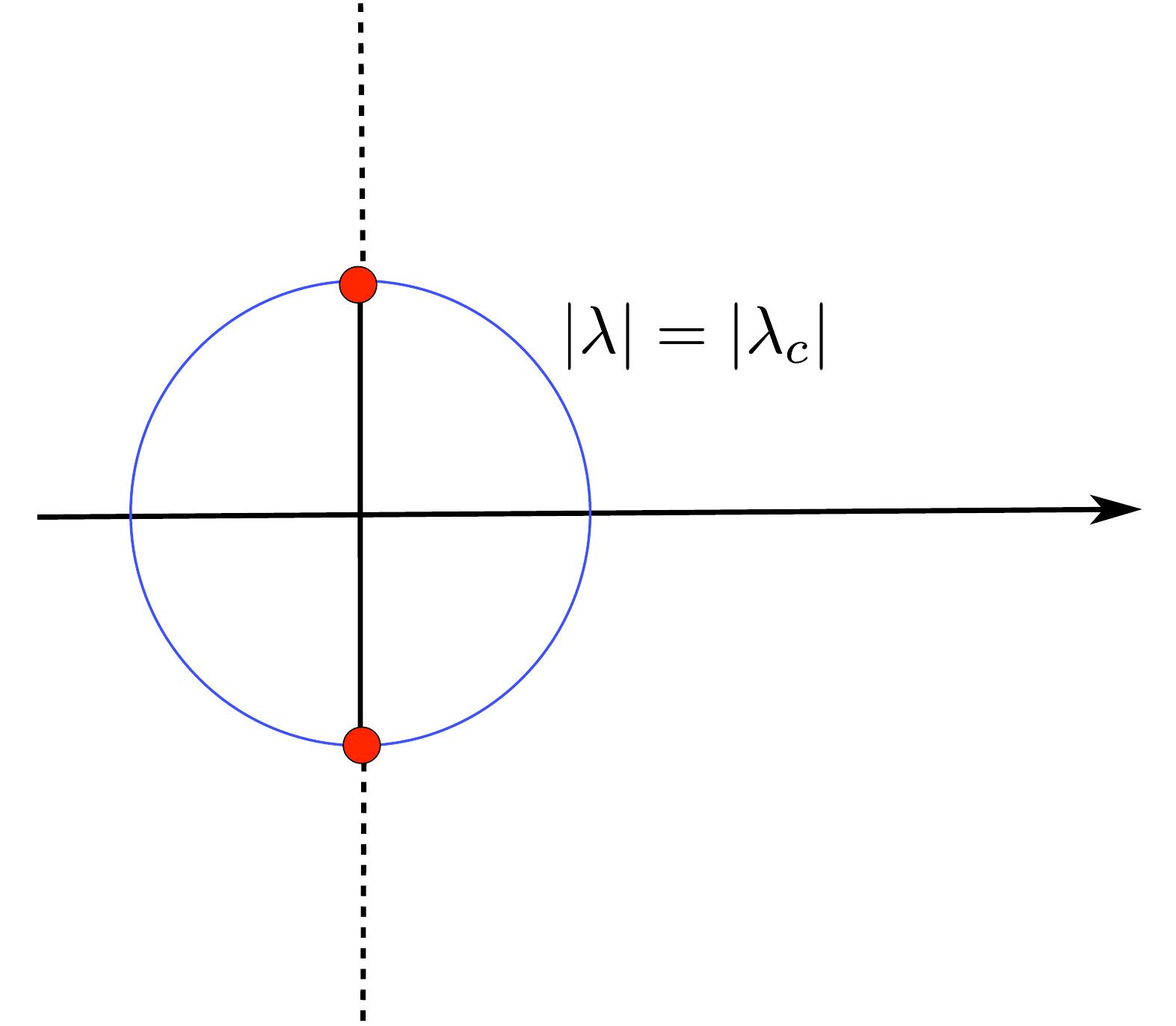}
\caption{The analyticity structure of the genus $g$ free energies leads to two branch cuts in the plane of the 't Hooft coupling, 
starting at the singularities $\pm \lambda_c$. The circle of radius $|\lambda_c|$ separates two different regions: 
there is a ``short distance phase" for $|\lambda|< |\lambda_c|$, and a ``long distance phase" for $|\lambda|> |\lambda_c|$.}
\label{thooftplane}
\end{figure}
One natural question which triggered this investigation is the following: the expansion (\ref{z-exp}), which is induced by the instanton expansion of the modified 
grand potential $J(\mu, k)$, is convergent for large values of $N$. Does it have a finite radius of convergence, i.e. is there a value of $N$ which signals a transition between 
two different regimes, which we could call a ``small distance" regime and a ``large distance" regime? Such a distinction between two different regimes can be made in the 
context of the 't Hooft expansion of $Z(N,k)$, which corresponds to the dual type IIA superstring. In the 't Hooft expansion, the natural quantities are the genus $g$ free energies, 
which are functions of the 't Hooft parameter $\lambda$. This is in turn related to the radius of the universe $L$ as 
\be
\left({L \over \ell_s} \right)^4 \approx \lambda, 
\ee
where $\ell_s$ is the string length. Therefore, $\lambda \approx 1$ corresponds to a stringy size universe. For $\lambda \gg 1$, we should have a geometric 
description in terms of embedded strings, while for small $\lambda$ the geometric description might break down. Indeed, as it can be seen from the explicit expression (\ref{standard}) for 
the genus zero free energy, there are branch cuts starting at 
\be
\kappa^2=-16, 
\ee
which corresponds to critical values of the 't Hooft parameter $\pm \lambda_c$, where \cite{dmpnp}
\be
\lambda_c= {2\ri K \over \pi^2}, 
\ee
and $K$ is Catalan's constant. In the $\lambda$ plane, the branch cuts are along 
the imaginary axis and they start at $\pm \lambda_c$, see \figref{thooftplane}. Although the genus $g$ free energies are smooth functions along the real axis, 
this non-trivial analytic structure leads to two well-defined regions in the complex $\lambda$-plane: a ``short distance phase" for $|\lambda|< |\lambda_c|$, 
and a ``long distance phase" for $|\lambda|> |\lambda_c|$. These 
two regions have a physical meaning, since for example the worldsheet instanton expansion of the planar free energy, which is convergent in the 
``long distance phase," will no longer converge in the ``short distance phase." At $|\lambda|= |\lambda_c|$ we then have a breakdown of the 
geometric description in terms of embedded strings. 
This picture of the moduli space of $\lambda$ is very similar to the picture developed for topological string theory  in \cite{agm} (see \cite{aspinwall} for 
an excellent summary of these developments). The non-trivial analytic structure found at the planar level is replicated in all the genus $g$ free energies, so 
we can say that, in this model, one sees two different ``phases" in the perturbative genus expansion. 

One interesting question is then: how does this non-trivial analytic structure get modified when we 
go to M-theory, i.e. when we consider $k$ fixed and we study the dependence on $N$ of $Z(N,k)$? 
The first issue we have to address is the following: in ABJM theory, $\lambda$ is necessarily a rational number, 
since it is the quotient of two integer numbers. However, the genus $g$ 
free energies obtained in the double-line expansion are already analytic functions in a neighborhood of the origin, in the complex $\lambda$ plane. One can 
then analytically continue them, and it makes sense to ask what is the analytic structure of the 
genus $g$ free energies obtained in this way, as functions of $\lambda$. In the same way, 
since $N$ is quantized, it does not make sense {\it a priori} to ask about the analytic properties of $Z(N,k)$, unless 
we find a way to promote $N$ to a full complex variable. However, this can be done in a natural way: the expression (\ref{naif}) gives $Z(N,k)$ as a Laplace transform of the 
modified grand potential. Therefore, we can use this expression to define $Z(N,k)$ for any complex value of $N$. 

This kind of promotion of a function defined only for positive integers, to a function defined on the complex plane, 
is similar to the promotion of the factorial to the Gamma function. It also happens in 
some simple matrix models. For example, the partition function of the Gaussian matrix model, which is defined in 
principle only for integer $N$, can be expressed for arbitrary $N$ in terms of the Barnes 
function $G_2(N+1)$ (see for example \cite{mmhouches}), therefore it can be extended to any complex $N$. The Borel resummation of the $1/N$ expansion 
also makes it possible to extend the definition of matrix model partition functions to continuous values of $N$ \cite{mmnp}. 
In our case, what makes it possible to perform this extension to non-integer values of $N$ 
is our underlying thermodynamic formalism. It is easy to check that this formalism can be used 
to obtain the standard promotion of the factorial to the Gamma function. Let us consider a simple statistical mechanical model: the classical ideal gas. 
After redefining the fugacity appropriately, the canonical partition function can be taken to be, 
\be
Z(N)={1\over N!}. 
\ee
The grand potential is,
\be
\CJ(\mu)=\re^\mu, 
\ee
and the inversion formula (\ref{finite-n}) reads, in terms of the fugacity (\ref{fugacity}), 
\be
\label{toy-model}
Z(N)= {1\over 2 \pi \ri} \oint_\CC \kappa^{-N-1} \re^\kappa \rd \kappa, 
\ee
where $\CC$ is a closed contour around the origin. We can now deform $\CC$ into a Hankel contour $\CH$ around the negative real axis, and the resulting integral can 
be used to define $Z(N)$ for any complex $N$: indeed, if $N$ is an integer, we simply recover (\ref{toy-model}) 
by contour deformation; if $N$ is not an integer, the integrand of (\ref{toy-model}) 
has a branch cut along the negative real axis, but since $\CH$ is a Hankel contour surrounding the cut, the integral is well defined. 
The resulting function is nothing but the Hankel representation of the 
reciprocal Gamma function (see for example \cite{henrici}, theorem 8.4b): 
\be
 {1\over 2 \pi \ri} \int_\CH \kappa^{-N-1} \re^\kappa \rd \kappa={1\over \Gamma(N+1)}.
 \ee

\begin{figure}
\center
\includegraphics[height=5cm]{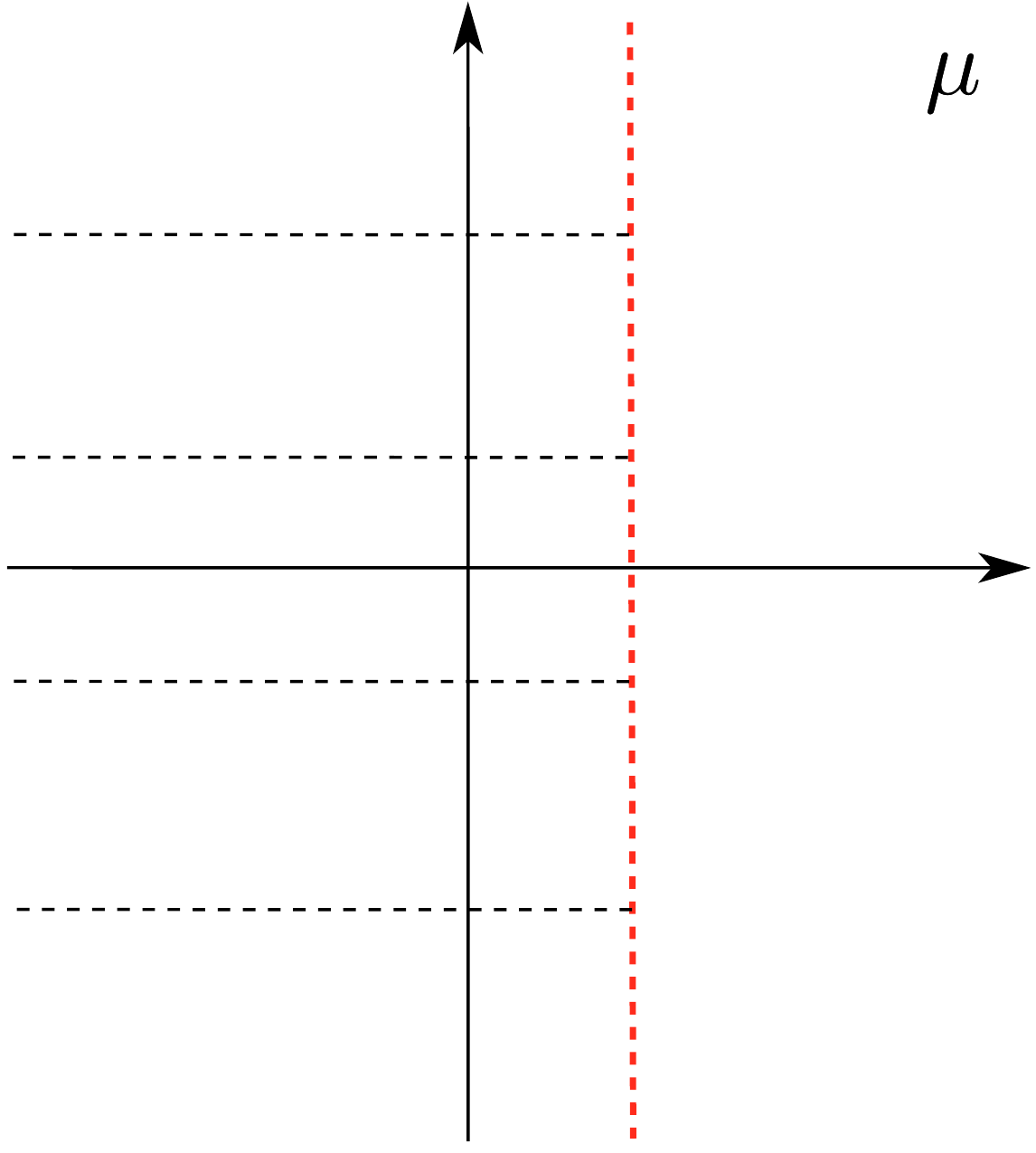}  \qquad \qquad  \qquad \includegraphics[height=5cm]{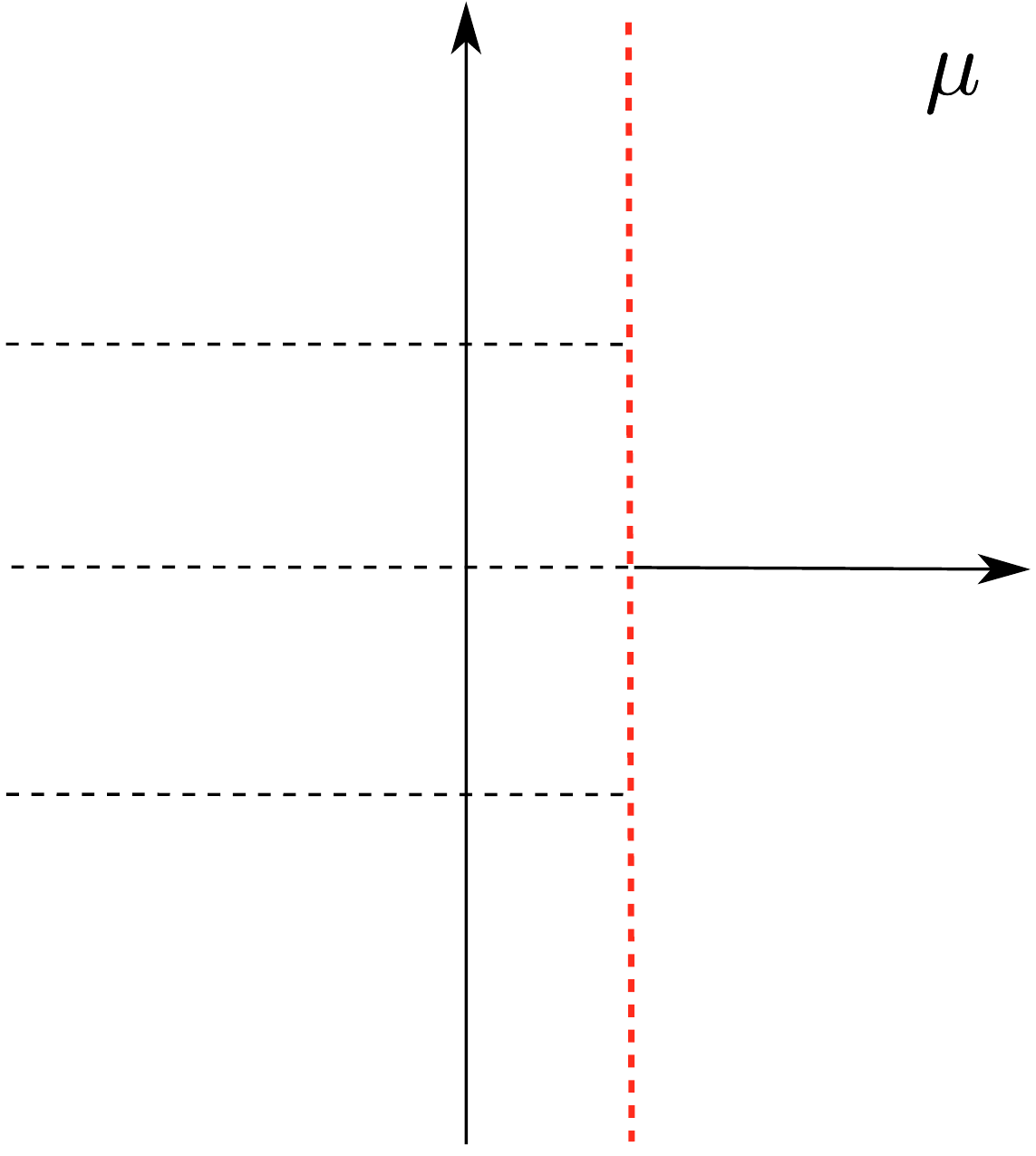}
\caption{The analyticity structure of the functions $J(\mu, 2)$ (left) and $J(\mu, 2;1)$ (right). Both functions are analytic for 
${\rm Re}(\mu) > 2 \log (2)$ (which is the region to the right of the vertical dashed line), but 
they have branch cuts when ${\rm Re}(\mu) < 2 \log (2)$, indicated here by horizontal dashed lines.}
\label{anal}
\end{figure}

Once that $Z(N,k)$ has been defined on the complex plane, we can ask what are its properties: when is it well-defined? Where is it analytic? 
To answer these questions, we first look at the analytic structure of the function appearing in the integrand, $J(\mu, k)$. This 
can be understood in detail in the cases with maximal supersymmetry, since in those cases we have completely explicit formulae for the 
modified grand potential. When $k=2$, 
the function $J(\mu,2)$ is analytic for $\mu > 2 \log (2)$, 
and it has branch cuts along the semi-infinite lines (see \figref{anal})
\be
{\rm Im}(\mu)=\left( {1\over 2} + n\right) \pi, \quad n \in \IZ, \qquad {\rm Re}(\mu)< 2 \log(2). 
\ee
The value $\mu=2 \log (2)$ signals the radius of convergence for the infinite series of non-perturbative effects entering into 
$J(\mu, 2)$ (i.e. this series is convergent for $\mu\in (2 \log(2), \infty)$.) 
For $k=1$ we have a very similar structure, but the radius of convergence is now $\mu=\log(2)$. 
Finally, for the ABJ theory with $M=1$ and $k=2$, the radius of convergence is again $\mu=2 \log(2)$, 
but the branch cuts are now at 
\be
{\rm Im}(\mu)=n \pi, \quad n \in \IZ, \qquad {\rm Re}(\mu)< 2 \log(2), 
\ee
as shown in \figref{anal}. Therefore, the analytic structure of the modified grand potential is similar to the one we found in the perturbative genus $g$ free energies: a non-trivial 
branch structure and a finite radius of convergence. 

\begin{figure}
\center
\includegraphics[height=6cm]{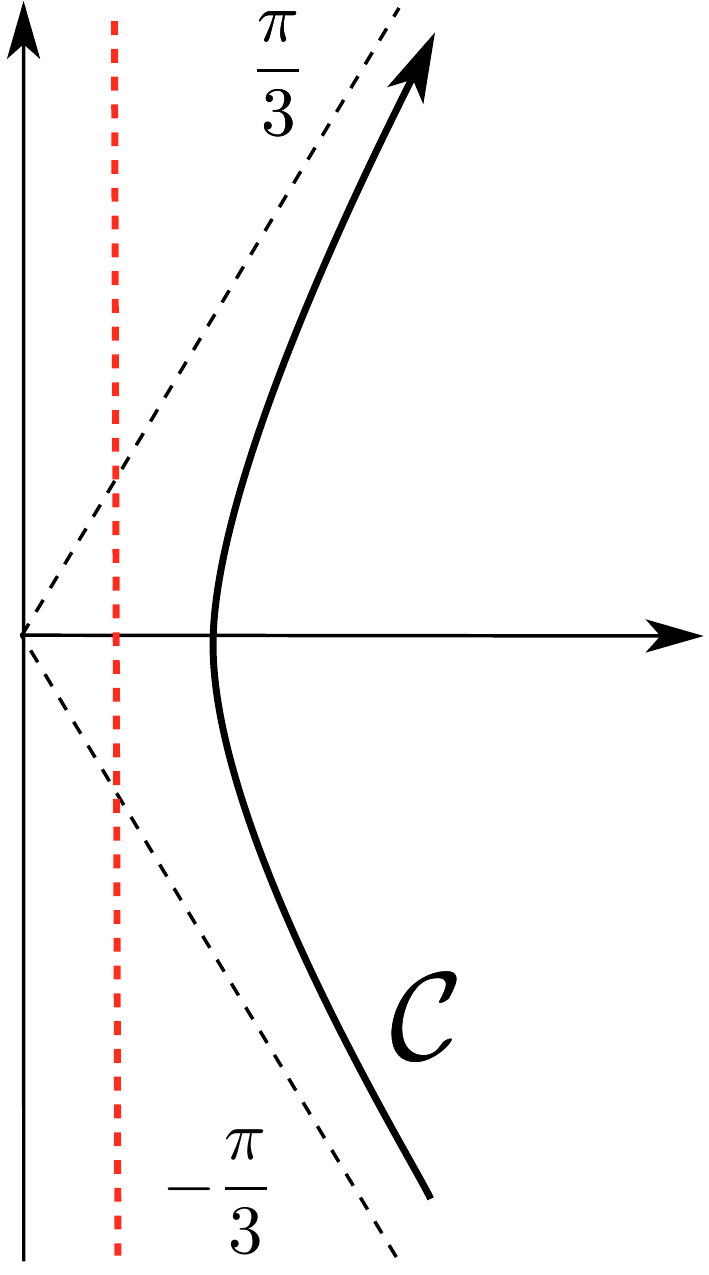}  
\caption{The contour of integration $\CC$ in (\ref{naif}) might be deformed to lie inside the region of analyticity of the modified grand potential, which is the semi-plane to the 
right of the vertical dashed line.}
\label{inside}
\end{figure}

The modified grand potential has in principle no direct physical meaning, and it 
should be regarded as an auxiliary object to define the partition function $Z(N,k)$ through the integral (\ref{naif}). To understand the properties of this integral we first 
note that the integration contour $\CC$ might be deformed so as to lie inside the region of analyticity of $J(\mu, k)$, for $k=1,2$, see \figref{inside}. By extending a 
standard result in the theory of Laplace transforms (see for 
example \cite{henrici}, chapter 10), we conclude that $Z(N,k)$, extended to a function on the complex plane of $N$ by (\ref{naif}) is in fact an {\it entire} function, 
for $k=1,2$. A similar argument holds of course for 
ABJ theory with $k=2$, $M=1$. 

\begin{figure}
\center
\includegraphics[height=5cm]{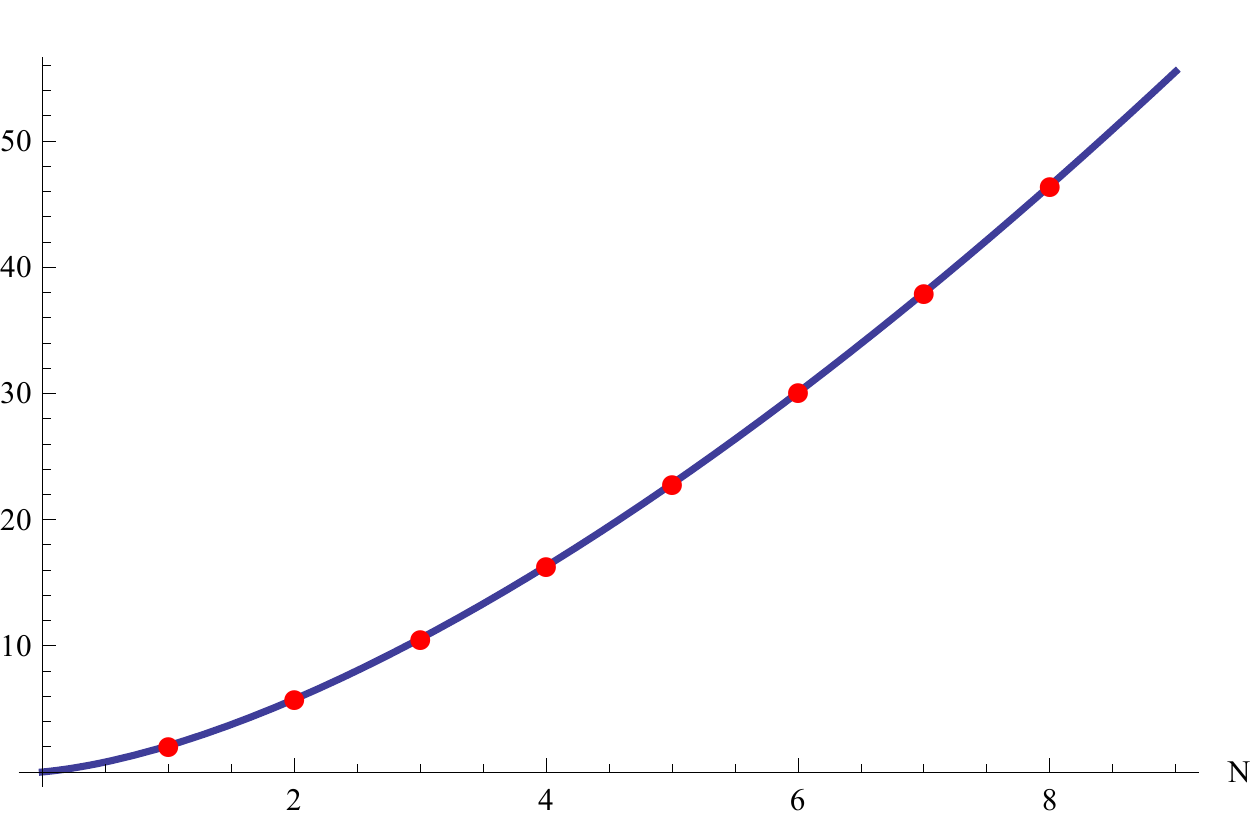}  
\caption{A plot of the interpolating function $-\log Z(N,2)$ as a function of real, positive $N$. The dots are the values obtained by evaluating the gauge theory partition function at the positive integers.}
\label{conti}
\end{figure}

We conclude that, in the maximally supersymmetric ABJ(M) theories, the partition functions $Z(N,k)$ and $Z(N,2;1)$, which are {\it a priori} only well defined 
for positive integer $N$, can be naturally extended to entire functions on the full complex plane. In \figref{conti}, we plot $-\log Z(N, 2)$ as a function of $N\ge 0$. The 
dots are the exact values of $-\log Z(N,2)$ for integer $N$, as obtained in for example \cite{hmo2}. We have then a natural function interpolating between 
the values at the positive integers, and extended analytically to the full complex plane. Note that one has $Z(0,k)=1$, for $k=1,2$, which is certainly natural from 
the point of view of the grand canonical partition function. 

 Since the partition function, as defined on the full complex plane, is analytic everywhere, 
 we do not have a natural way to distinguish a short-distance phase 
 from a long-distance phase, since the regime of small $N$ is continuously connected to the regime of large $N$ through the entire function $Z(N,k)$. 
 It is tempting to interpret this result by saying that, as we go to M-theory, we ``erase" the non-trivial analytic structure found in 
 the perturbative genus expansion, and we obtain a much simpler quantum geometry for the moduli space, which is now just the complex plane. 
 This is similar to what was found in \cite{mmss} in the context of minimal strings, 
where it was shown that the moduli space of a FZZT brane, which has a non-trivial 
analytic structure in perturbation theory, becomes just the complex plane in the exact theory.  

In the case of ABJM theory with general $k$, we do not have analytic expressions for the modified grand potential. 
However, one can compute the series of non-perturbative corrections to relatively large order, 
and evaluate the resulting (approximate) partition function from (\ref{naif}). The results we have obtained indicate that, as in the maximally supersymmetric case, the contour 
$\CC$ falls inside the region of analyticity of $J(\mu, k)$ and one obtains an entire function $Z(N,k)$ on the full complex plane. 

The existence of this interpolating function raises interesting conceptual issues. Clearly, such a continuation is not unique. However, is there any sense in which (\ref{naif}) is a preferred choice for 
this function? Does the existence of this function indicate that we can go beyond the discretization of the geometry, which comes from the realization of the geometry in terms of 
gauge theory, and therefore seems to be built in the AdS/CFT correspondence? 

\subsection{A de Sitter continuation?} 

\begin{figure}
\center
\includegraphics[height=5cm]{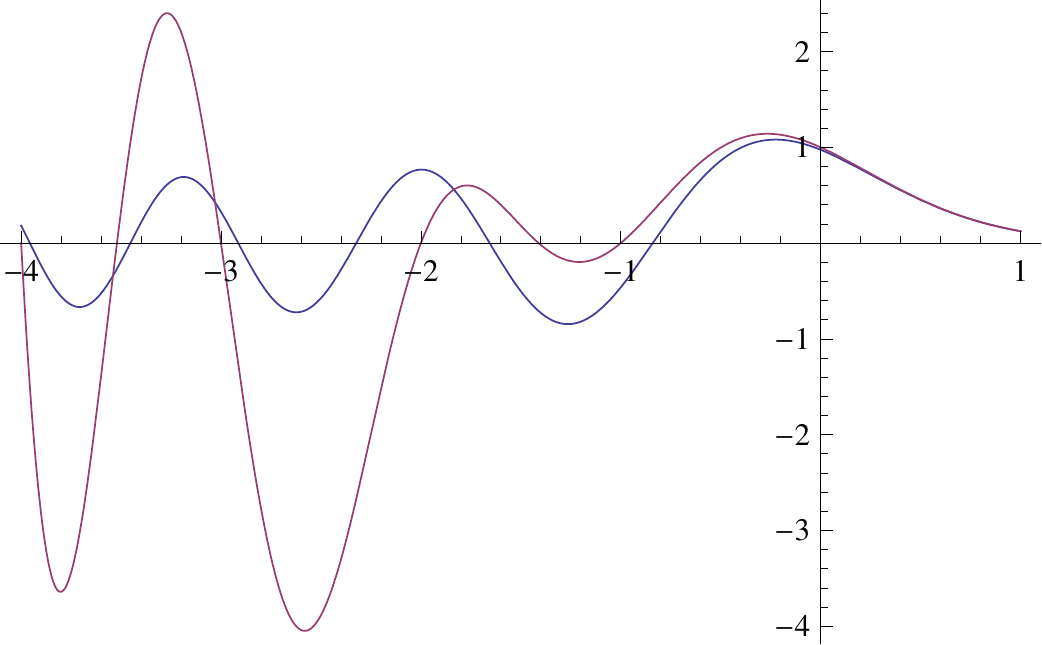}  
\caption{Here we show, as a function of $N$, the function $Z(N,2)$, as well as its perturbative piece, which is essentially an Airy function. 
For positive $N$, the exact result is well approximated by its perturbative limit. However, 
when $N$ is negative, the non-perturbative corrections take over and lead to wide fluctuations in $Z(N,2)$.}
\label{desitter}
\end{figure}

It has been suggested that many results in AdS/CFT can be extended to de Sitter space,  
by simply performing an analytic continuation of the radius $L \rightarrow \ri L$ 
\cite{malda}. After this continuation, the partition function of the Euclidean CFT is interpreted as 
a wavefunction in de Sitter space. In the context of ABJM theory in the M-theory regime (i.e. for $k$ fixed), this continuation 
corresponds to switching $N \rightarrow -N$, as it follows from the dictionary (\ref{ads-dictio}). 
Motivated by recent proposals and calculations along these lines 
(see \cite{ahs,ad,castrom}, for recent examples), and since we now have a natural extension of 
$Z(N,k)$ to the full complex plane, we can look at the behavior of this partition function for negative $N$. As shown in \figref{desitter}, one finds an oscillatory 
function which seems to have zeros at the negative integers (just like $1/\Gamma(N+1)$). 

An interesting aspect of the behavior at negative $N$ is the following: the function $Z(N,k)$ is given by a perturbative piece, 
which is essentially an Airy function, together with an infinite series of 
corrections coming from non-perturbative effects, as in (\ref{leadingco}). When $N>0$, these corrections are exponentially suppressed and small. 
However, for negative $N$, they become important. 
This is of course due to the fact that they go like $\exp (- {\sqrt{N}})$, hence they are no longer suppressed for $N<0$ and lead to oscillatory factors. The resulting behavior
is shown in \figref{desitter}, where we plot both the exact $Z(N,2)$ and its perturbative piece. The moral of this story is that an approximation 
which was reasonable for for $N>0$ (obtained in this case by dropping the contribution of non-perturbative effects) is no longer reasonable once we go to 
negative $N$. This might be a warning for 
calculations in which one starts with a truncated answer in Euclidean AdS, which is then continued to de Sitter space: the corrections which have not 
been kept might become crucial after the analytic continuation. It might be also a warning for the non-perturbative consistency 
of the continuation of Euclidean AdS results to de Sitter (see \cite{belin} for a discussion of related issues.)

\sectiono{Conclusions}

In this paper we have shown that non-perturbative effects in the partition function of ABJ(M) theories simplify drastically in the maximally supersymmetric case. As a 
consequence, one can write down closed formulae for the grand potential introduced in \cite{mp,hmo2}, which encodes the full large $N$ asymptotics. This makes it possible to perform precision tests 
of the proposals of \cite{mp,hmo2, hmo3,hmmo,km} and their generalizations to ABJ theory \cite{matsumori,hondao,kallen}. More surprisingly, we have been able to write explicit formulae 
for the grand canonical partition functions, which give generating functionals for the partition functions and involve the formalism of \cite{bde,eynard, em}. The exact 
quantization conditions conjectured in \cite{km,kallen} have then a beautiful realization as zeroes of Jacobi theta functions. 
Clearly, it would be very interesting to understand better these results, 
to extend them to other models, and/or to derive them rigorously. In particular, it would be interesting to see 
whether the grand canonical partition function of the $\CN=6$ theories (where no truncation 
of the free energies to $g=0,1$ takes place) can be related to the full formalism of \cite{eynard, em}, incorporating all the $1/N$ corrections.

We have used our exact results to explore some aspects of the quantum geometry of M-theory. We have seen that the 
analytic properties found in the M-theory context turn out to be quite 
different from the ones found in the type IIA superstring. In order to perform this comparison, we promoted 
the partition function of ABJ(M) theory, which is in principle only 
defined for discrete values of the rank of the gauge group, to an entire function on the complex plane. We have also 
explored some properties 
of this extension for negative values of $N$, since this has been suggested to be related to the physics of de Sitter space. 
The existence of this function also raises interesting conceptual issues touching on the 
foundations of the AdS/CFT correspondence: 
can we define non-perturbative objects in gauge theory for arbitrary values of $N$? 
Is this definition unique, or is there a natural way to single out a preferred choice? 
What would be the physical meaning of such definitions? What would be their implications for the 
short-distance regime of gravity theories? We hope that the exact results presented in this paper 
will be helpful in future investigations of these issues.

\section*{Acknowledgements}
We would like to thank Alexandre Belin, Stefano Cremonesi, Rajesh Gopakumar, Yasuyuki Hatsuda, Elias Kiritsis, Igor Klebanov,
Bruno Le Floch, Domenico Orlando, Boris Pioline, Susanne Reffert and Andy Strominger 
for useful discussions and correspondence. S.C., A.G. and M.M. would like to thank the organizers of the Ascona conference on ``Recent 
developments in string theory," for providing a stimulating atmosphere. M.M. would like to thank CERN, 
and also the CERN-Korea Theory Collaboration funded by National Research Foundation (Korea),
for giving him the opportunity to present this work in the conference ``Exact results in SUSY gauge theories in various dimensions." He would 
also like to thank the organizers and the participants in this conference 
for their comments and constructive criticism. 
This work is supported in part by the Fonds National Suisse, 
subsidies 200020-137523 and 200020-141329, and by the NCCR SwissMAP. 

\appendix 

\sectiono{The special geometry of local $\IP^1 \times \IP^1$}

In this Appendix we collect some useful results on the special geometry of local $\IP^1 \times \IP^1$. A summary can be found in \cite{dmp,hmmo}. 

In principle, the special geometry of local $\IP^1 \times \IP^1$ involves a moduli space of complex dimension $2$, which can be parametrized by the ``bare" 
parameters $z_1, z_2$. In this full two-parameter model, the periods are given by the following formulae (we changed the signs of the bare parameters in \cite{hmmo} in order to 
obtain the correct formula in ABJM theory with $k=2$). The $A$-periods are given by 
\be
\Pi_{A_I}(z) =\log z_I + \widetilde \Pi_A (z_1, z_2), \qquad I=1,2,
\ee
where
\be
\label{Aper}
\widetilde \Pi_A (z_1, z_2)= 2\sum_{k,l\ge 0, \atop (k,l)\not=(0,0)} { \Gamma(2k + 2l) \over \Gamma(1+k)^2 \Gamma(1+l)^2} (-z_1)^k (-z_2)^l .\ee
There are two independent $B$-periods, $\Pi_{B_I}(z_1, z_2)$, $I=1,2$, which are related by the exchange of $z_1$ and $z_2$, 
\be
\Pi_{B_2}(z_1, z_2)=\Pi_{B_1}(z_2, z_1). 
\ee
The $B_1$ period is given by 
\be
\label{B1per}
\Pi_{B_1}(z_1, z_2)=-{1\over 8}\left( \log^2 z_1  -2 \log z_1\log z_2 -\log^2 z_2 \right) +{1\over 2} \log z_2\,  \widetilde \Pi_A (z_1, z_2)+ {1\over 4}  \widetilde \Pi_B (z_1, z_2) 
\ee
where
\be
\label{Bper}
\ba
\widetilde \Pi_B (z_1, z_2) &= 8 \sum_{k,l\ge 0, \atop (k,l)\not=(0,0)} { \Gamma(2k + 2l) \over \Gamma(1+k)^2 \Gamma(1+l)^2} \left( \psi(2k+2l) -\psi(1+l)\right) (-z_1)^k (-z_2)^l 
 . \ea
 \ee
 This determines the standard prepotential by 
 \be
 \partial_{t_1} F_0(t_1, t_2)= \Pi_{B_1}(z_1, z_2), 
 \ee
 together with the requirement of symmetry. 
 
 The restriction to the diagonal theory $z_1=z_2$ gives the two periods
 \be
 \label{vares}
 \ba
 \varpi_1(z)&=\log z + \widetilde \varpi_1(z), \\
 \varpi_2(z)&=\log^2(z) +2 \log z  \widetilde \varpi_1(z) + \widetilde \varpi_2(z),
 \ea
 \ee
 where
\be
\ba
\widetilde \varpi_1 (z)&=\widetilde \Pi_{A}(z,z)=\sum_{n \ge 1}  {1\over  n} \left( {\Gamma \left( n+{1\over 2} \right) \over 
\Gamma({1\over 2}) n!} \right)^2 16^n (-z)^n, \\
\widetilde \varpi_2 (z)&=\widetilde \Pi_{B}(z,z)\\
&=\sum_{n \ge 1}   {4 \over  n} \left( {\Gamma \left( n+{1\over 2} \right) \over 
\Gamma({1\over 2}) n!} \right)^2 16^n \left[ \psi\left(n+{1\over 2} \right) -\psi(n+1)+ 2 \log 2-{1\over 2n} \right] (-z)^n. 
\ea
\ee
The periods $\varpi_{1,2}(z)$ are the two fundamental solutions to the Picard--Fuchs equation $\CL \Pi=0$, where \cite{kz}
\be
\label{pf-eq}
\CL=\theta^3 - 4z \theta (2 \theta+1)^2.
\ee
They can be written in closed form in terms of special functions. We have, 
\be
\label{tildo}
\widetilde \varpi_1 (z)=-4 z \, _4F_3\left(1,1,\frac{3}{2},\frac{3}{2};2,2,2;-16 z\right), 
\ee
as well as
\be
\ba
\widetilde \varpi_2(z)& =-\log^2(z) -{\pi^2 \over 3} -2 \ri \pi  \log (z)+8 \ri \pi  z \, _4F_3\left(1,1,\frac{3}{2},\frac{3}{2};2,2,2;-16 z\right)\\
&+8
   z \log (z) \, _4F_3\left(1,1,\frac{3}{2},\frac{3}{2};2,2,2;-16
   z\right)+\frac{2}{\pi } G_{3,3}^{3,2}\left(-16 z\left|
\begin{array}{c}
 \frac{1}{2},\frac{1}{2},1 \\
 0,0,0
\end{array}
\right.\right), 
\ea
\ee
which means that
\be
\varpi_2(z)= \frac{2}{\pi } G_{3,3}^{3,2}\left(-16 z\left|
\begin{array}{c}
 \frac{1}{2},\frac{1}{2},1 \\
 0,0,0
\end{array}
\right.\right)- 2 \pi \ri \varpi_1(z) -{\pi^2 \over 3}. 
\ee
The mirror map is 
\be
Q= \re^{-t}=z \exp \left( \widetilde \varpi_1 (z)\right), 
\ee
or, equivalently, 
\be
\label{mirror-map}
t=-\varpi_1(z)=-\log z + 4 z \, _4F_3\left(1,1,\frac{3}{2},\frac{3}{2};2,2,2;-16 z\right).
\ee
With our conventions, the genus zero free energy or prepotential of the diagonal theory has the following expansion at large radius, 
\be
F_0(t)= {t^3 \over 6} -2 \zeta(3) + 4 Q -{ 9Q^2 \over 2}  + {328 Q^3 \over 27} -{777 Q^4 \over 16}+\CO(Q^5). 
\ee
The derivatives of the periods are hypergeometric functions, which in this case can be expressed in terms of 
elliptic integrals of the first kind. We have
\be
{z {\rd \over \rd z}}  \widetilde \varpi_1 (z)= -1+ {2 K(-16 z) \over \pi}. 
\ee
as well as 
\be
z{\rd \over \rd z}G_{3,3}^{3,2}\left(-16 z\left|
\begin{array}{c}
 \frac{1}{2},\frac{1}{2},1 \\
 0,0,0
\end{array}
\right.\right)=-2 \pi K' (-16z)=-2 \pi K(1+16z).
\ee
Here, the argument of the elliptic integrals is the squared modulus $k^2$. It follows that
\be
\label{ddtf}
\partial_t^2 F_0 (t)=\pi {K (1+ 16 z) \over K(-16 z)} +\pi \ri. 
\ee

Using these and other results in special geometry, one can derive the identity (\ref{off-no}). This goes as follows. 
First, we take a derivative of the l.h.s. of (\ref{off-no}) w.r.t. $t$. This can be expressed in terms of Yukawa couplings 
(w.r.t. the flat coordinates)
\be
\partial_t \left( \partial_{t_1} -\partial_{t_2}\right)^2 F_0(t_1, t_2) \big|_{t_1=t_2=t} =2\left( C_{t_1 t_1 t_1} - C_{t_1 t_1 t_2} \right)\big|_{t_1=t_2=t}. 
\ee
In order to compute the r.h.s., we also need the Jacobian of the mirror map $\partial z_i /\partial t_j$, restricted to the diagonal. It is easy to see, form the explicit expression 
of the mirror map, that
\be
\ba
{\partial z_1 \over \partial t_1} \bigg|_{t_1=t_2=t} &={\partial z_2 \over \partial t_2} \bigg|_{t_1=t_2=t}  = {z\over 4} \left( 2+ {\pi \over K(-16 z)} \right), \\
{\partial z_1 \over \partial t_2} \bigg|_{t_1=t_2=t} &={\partial z_2 \over \partial t_1} \bigg|_{t_1=t_2=t}  = {z\over 4} \left( -2+ {\pi \over K(-16 z)} \right).
\ea
\ee
By using the explicit expression for the Yukawas presented in for example \cite{dmp} (and after changing $z_i \rightarrow -z_i$), we find, in terms of $z$, 
\be
\partial_t \left( \partial_{t_1} -\partial_{t_2}\right)^2 F^{\rm inst}_0(t_1, t_2) \big|_{t_1=t_2=t} = 1-{\pi \over 2 K(-16 z)}, 
\ee
and by integrating once w.r.t. $t$ we obtain (\ref{off-no}). 

In going from the large radius frame to the orbifold frame we have to express  the large radius periods in terms of  the orbifold periods 
(\ref{standard}). This is done by using the following relations
\be 
\label{per-anc} \ba t &={1\over \pi^2}\partial_{\lambda}\CF_0, \\
G_{3,3}^{3,2}\left(-16 z|
\begin{array}{c}
 \frac{1}{2},\frac{1}{2},1 \\
 0,0,0 \\
\end{array}
\right)&=4 \pi^3 \lambda -\ri \partial_{\lambda}\CF_0.
\ea\ee

\sectiono{Jacobi theta functions}

Our conventions for the Jacobi theta functions are as in \cite{akhiezer},
\be
\label{jtheta}
\ba
\vartheta_1(v,\tau)&= \sum_{n \in \IZ} (-1)^{n-1/2} \exp \left[ \pi \ri \left( n+1/2\right)^2 \tau + 2 \pi \ri \left( n +1/2\right) v\right], \\
\vartheta_2(v,\tau)&= \sum_{n \in \IZ}  \exp \left[ \pi \ri \left( n+1/2\right)^2 \tau  + 2 \pi \ri \left( n +1/2\right) v\right], \\
\vartheta_3(v,\tau)&= \sum_{n \in \IZ}  \exp \left[ \pi \ri n^2 \tau + 2 \pi \ri n v\right], \\
\vartheta_4 (v,\tau)&=\sum_{n \in \IZ}  (-1)^n \exp \left[ \pi \ri n^2 \tau + 2 \pi \ri n v\right]. 
\ea
\ee
They satisfy the following properties, 
\be
\label{theta-ids}
\ba
 \vartheta_3(v, \tau)&=\vartheta_3(2v, 4 \tau)+\vartheta_2(2v, 4 \tau),\\
 \vartheta_3(v, \tau)&=\vartheta_4\left(v+1/2, \tau \right),\\ 
  \vartheta_2(v, \tau)&=\vartheta_1\left(v+1/2, \tau \right),
  \ea
  \ee
 and they transform under an $S$-transformation as, 
\be
\label{mod-trans}
\ba
\vartheta_3(v, \tau)={1\over \alpha}\vartheta_3\left(v/\tau, -1/\tau\right),\\
 \vartheta_4(v, \tau)={1\over \alpha}\vartheta_2\left(v/\tau, -1/\tau\right),\\
 \vartheta_1(v, \tau)={\ri \over \alpha} \vartheta_1\left(v/\tau, -1/\tau\right),
 \ea
 \ee
where 
\be 
 \alpha=\left(- \ri \tau \right)^{1\over 2}\re^{\pi \ri v^2 /\tau}. 
 \ee

\end{document}